\documentclass[aps,prl,superscriptaddress,twocolumn]{revtex4}
\usepackage{graphicx}
\usepackage{amsmath}
\usepackage{bm}
\usepackage{pdfpages}
\usepackage{stmaryrd}
\usepackage{mathrsfs}

\usepackage{mathrsfs}

\usepackage{SIunits}
\begin{document}

\title{Dirac, Rashba and Weyl type spin-orbit couplings: toward experimental realization in ultracold atoms}

\author{Bao-Zong Wang}
\affiliation{International Center for Quantum Materials, School of Physics, Peking
University, Beijing, 100871, China}
\affiliation{Collaborative Innovation Center of Quantum Matter, Beijing 100871,
China}
\affiliation{Shanghai Branch, National Laboratory for Physical Sciences
at Microscale and Department of Modern Physics, University
of Science and Technology of China, Shanghai 201315,
China}

\author{Yue-Hui Lu}
\affiliation{International Center for Quantum Materials, School of Physics, Peking
University, Beijing, 100871, China}
\affiliation{Collaborative Innovation Center of Quantum Matter, Beijing 100871,
China}

\author{Wei Sun}
\affiliation{Shanghai Branch, National Laboratory for Physical Sciences
at Microscale and Department of Modern Physics, University
of Science and Technology of China, Shanghai 201315,
China}
\affiliation{Chinese Academy of Sciences (CAS) Center for
Excellence and Synergetic Innovation Center of Quantum
Information and Quantum Physics, University of Science and
Technology of China, Hefei, Anhui 230026, China}

\author{Shuai Chen}
\thanks{shuai@ustc.edu.cn}
\affiliation{Shanghai Branch, National Laboratory for Physical Sciences
at Microscale and Department of Modern Physics, University
of Science and Technology of China, Shanghai 201315,
China}
\affiliation{Chinese Academy of Sciences (CAS) Center for
Excellence and Synergetic Innovation Center of Quantum
Information and Quantum Physics, University of Science and
Technology of China, Hefei, Anhui 230026, China}

\author{Youjin Deng}
\affiliation{Shanghai Branch, National Laboratory for Physical Sciences
at Microscale and Department of Modern Physics, University
of Science and Technology of China, Shanghai 201315,
China}
\affiliation{Chinese Academy of Sciences (CAS) Center for
Excellence and Synergetic Innovation Center of Quantum
Information and Quantum Physics, University of Science and
Technology of China, Hefei, Anhui 230026, China}

\author{Xiong-Jun Liu}
\thanks{xiongjunliu@pku.edu.cn}
\affiliation{International Center for Quantum Materials, School of Physics, Peking
University, Beijing, 100871, China}
\affiliation{Collaborative Innovation Center of Quantum Matter, Beijing 100871,
China}



\begin{abstract}
We propose a hierarchy set of minimal optical Raman lattice schemes toward the experimental realization of spin-orbit (SO) couplings of various types for ultracold atoms, including two-dimensional (2D) Dirac type of $C_4$ symmetry, 2D Rashba and 3D Weyl types.
These schemes are well accessible with current cold-atom technology, and in particular, a long-lived Bose-Einstein condensation of the 2D Dirac SO coupling has been experimentally proved. The generation of 2D Rashba and 3D Weyl types has an exquisite request that two sources of laser beams have distinct frequencies of factor-two difference. Surprisingly, we find that $^{133}$Cs atoms provide an ideal candidate for the realization.
A common and essential feature is the absence of any fine-tuning and phase-locking in the realization, and the resulted SO coupled ultracold atoms have a long lifetime. These schemes essentially improve over the current experimental accessibility and controllability,
and open new experimental platforms to study high-dimensional SO physics and novel topological phases with ultracold atoms.
\end{abstract}

\maketitle

Topological phase of matter has become a mainstream of research in condensed matter physics.
Recent outstanding examples include topological insulators,
which have been predicted and experimentally discovered in two-dimensional (2D) and 3D materials~\cite{hasan2010colloquium,qi2011topological},
topological semimetals, which exhibit linear dispersion around nodes termed the Dirac or Weyl points,
were found recently in a number of materials like Cd$_3$As$_2$~\cite{wang2013three,liu2014stable} and TaAs~\cite{xu2015discovery,lv2015experimental},
and topological superconductors~\cite{wilczek2009majorana,alicea2012new,Franz2013a,Franz2015RevModPhys},
which host exotic zero-energy states called Majorana modes and have attracted considerable experimental efforts~\cite{Mourik1003,Deng2012,Das2012,Perge2014,Ruby2015,Meyer2016,JiaJin-Feng2016}.
In these matters, the spin-orbit (SO) interactions, manifesting Dirac, Rashba and  Weyl types, play essential roles in driving the phases to be topologically nontrivial.

Very recently, the experimental realizations of 2D SO couplings have been respectively reported in $^{87}$Rb Bose-Einstein condensate (BEC)~\cite{wu2016realization} and
$^{40}$K degenerate Fermi gases~\cite{huang2016experimental,meng2016experimental}, where nontrivial band topology and Dirac point are observed.
The reports signify an important achievement in the field of quantum simulator with ultracold atoms and optical lattice, opening a great deal of possibilities to explore novel quantum states beyond natural conditions~\cite{zhang2008p,sato2009non,goldman2010realistic,sau2011chiral,liu2013detecting,cui2014universal,liu2014realizationtogether,zhou2014three,goldman2014light,xu2015berezinskii,chan2017non,zhu2017exploring,song2017observation,gong2017zeno,chan2017generic}.
Nevertheless, the current experiments suffer challenges in controllability, since the realizations rely on the blue-detuned optical lattices~\cite{wu2016realization,liu2014realizationtogether} or delicate manipulation of resonant Raman couplings between multiple ground states~\cite{huang2016experimental,meng2016experimental}. To be broadly applicable to studying exotic topological phases mentioned above,
the SO coupled systems should exhibit the following criteria, the high controllability and high stability with long lifetime,
which however cannot be satisfied in the current experiments.

\begin{figure}
\includegraphics[scale=0.3]{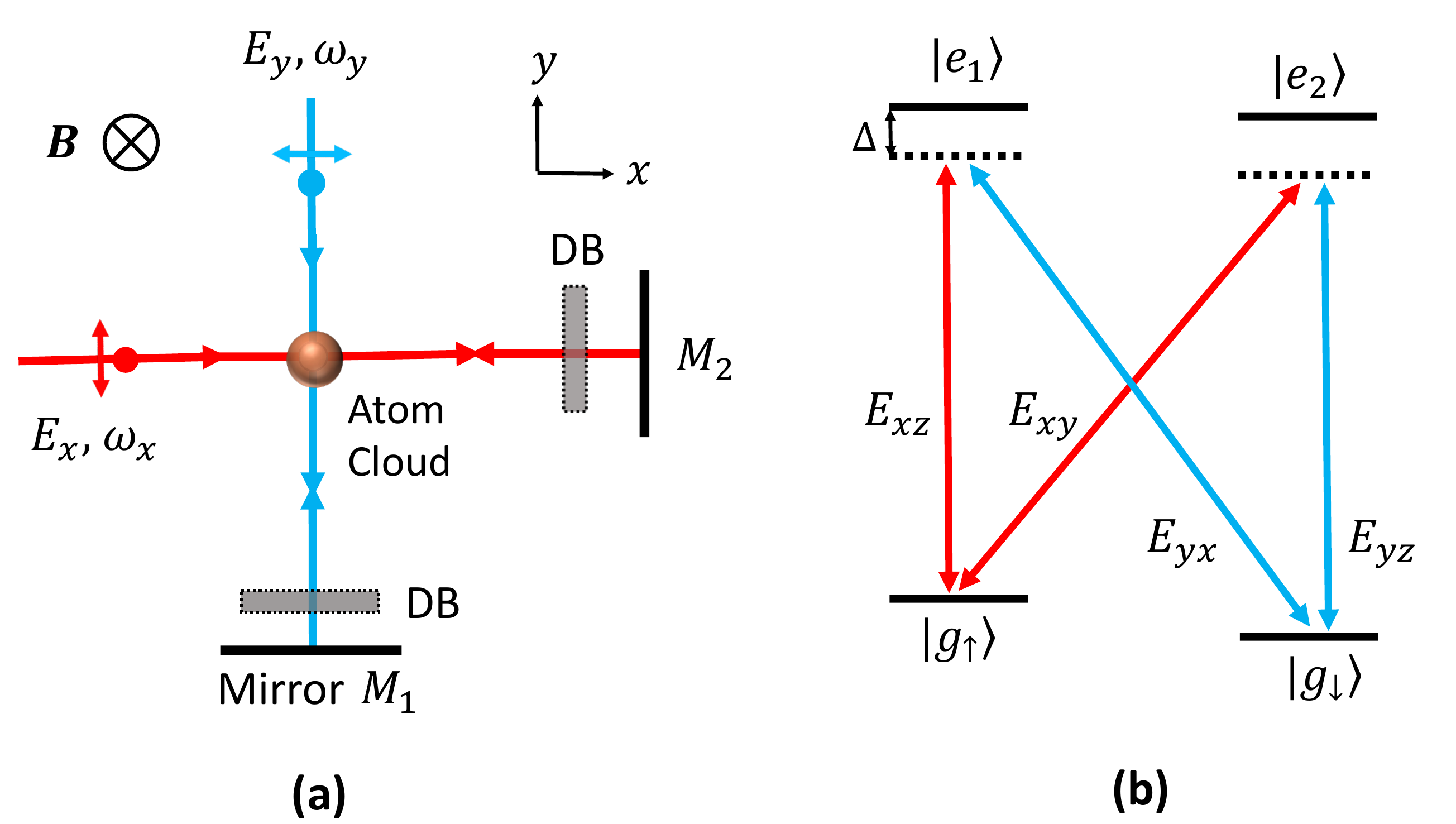}
\caption{(a) Generic setup for realization of Dirac type 2D SO coupling.
A pair of laser beams, $\bm{E}_x$ and $\bm{E}_y$, produce square lattice and two independent Raman potentials.
The dotted boxes, denoted as DB, contain $\lambda/4$ wave plates used to manipulate the symmetry of the system.
(b) The two Raman potentials are generated through the $double \text{-} \Lambda$ type configuration, with the detuning $\Delta$ being red or blue.}
\label{Fig:realization_Dirac}
\end{figure}

In this Letter, we propose a hierarchy set of minimal optical Raman lattice schemes to realize high-dimensional SO couplings of various types in cold atoms,
including 2D Dirac,  Rashba and 3D Weyl types.
The schemes make use of spin-independent or spin-dependent lattice and periodic Raman potentials,
the latter of which are generated via a $double \text{-} \Lambda$ internal level configuration.
For the Dirac SO coupling, the current new scheme has essential advantages in two primary aspects. First, the new realization is of high controllability, well suited for both red- and blue-detuned lattices.
Further, it exhibits a precise $C_4$ symmetry, leading to much broader topological phase region and high stability with a lifetime one-order over that in our previous realization~\cite{wu2016realization}.
These remarkable features are confirmed by the successful generation of a long-lived 2D Dirac SO coupling for $^{87}$Rb BEC~\cite{experimentalpaper}.
Realization of the Rashba and Weyl type SO couplings applies a spin-dependent lattice, and
it further requests that the Raman-coupling beams have a frequency which is double of that for the lattice beam.
This seemingly stringent requirement can be surprisingly well satisfied by $^{133}$Cs atoms which provide an ideal candidate for the present scheme.
A long lifetime over several seconds is also predicted for the realization with $^{133}$Cs atoms, which shows high feasibility of the scheme for Rashba and Weyl SO couplings.

\textit{Dirac type 2D SO coupling with $C_4$ symmetry.}
The realization is sketched in Fig.~\ref{Fig:realization_Dirac}(a).
The basic elements include a pair of laser beams which generate both the optical lattice and the Raman couplings,
with two sets of ``dotted boxes" (denoted as DB) containing $\lambda/4$ wave plates used to manipulate the symmetry of the realized Hamiltonian.
The spin states $|g_{\uparrow,\downarrow}\rangle$ refer to as two hyperfine levels, which are split by a bias magnetic field along the $\hat{z}$ direction.
The laser beam $\bm{E}_x$ running in the $x$ direction has frequency $\omega_x$ and polarization in the $y \text{-} z$ plane,
and $\bm{E}_y$ in the $y$ direction has frequency $\omega_y$ and polarization in the $x \text{-} z$ plane, with their wave vector $k_0 = \omega_x / c\approx\omega_y/c$.
The two beams generate both the lattice and Raman potentials when the frequency difference $\delta \omega = \omega_y - \omega_x$ compensates the Zeeman splitting between $|g_{\uparrow}\rangle$ and $|g_\downarrow\rangle$.
We show below that this minimized simple setting generates the following Hamiltonian~\cite{SI}
\begin{align}
H&= \left[ \frac{\bm{p}^2}{2m} + V_{\rm latt} (x,y) \right] \otimes \bm{1}  + m_z \sigma_z \nonumber  \\
 &{} \ + \mathcal{M}_x (x,y) \sigma_x + \mathcal{M}_y (x,y) \sigma_y \; ,
 \label{eq:Hamiltonian}
\end{align}
where $V_{\rm latt}$ denotes the square lattice potential, $\mathcal{M}_{x/y}$ are the Raman coupling potentials,
and $m_z = \delta/2$ measures the two-photon detuning $\delta$ of Raman coupling.

The ``DB" box containing a $\lambda/4$ wave plate induces an additional $\pi/2$-phase shift for $\bm{\hat{z}}$-component field.
This gives the light fields as $\bm{E}_x =  \bm{\hat{y}} E_{xy} \cos k_0x + i \bm{\hat{z}} E_{xz} \sin k_0x$,
and $\bm{E}_y = \bm{\hat{x}} E_{yx} \cos k_0y + i \bm{\hat{z}} E_{yz} \sin k_0y$, where $E_{\mu \nu} (\mu, \nu = x,y,z)$ is
amplitude of the field in the $\mu$ direction and with the $\nu$ polarization. All the irrelevant phase factors in $\bm{E}_{x,y}$ have been ignored.
For alkali atoms, the scalar optical potential generated by linearly polarized lights is spin-independent for typical detuning $\Delta$ which is less than fine structure splitting but much larger than hyperfine structure splitting. It follows that $V_{\rm latt} (x,y) = V_{0x} \cos^2 k_0x + V_{0y} \cos^2 k_0y$, with the amplitudes $V_{0x}\propto ( E_{xy}^2 - E_{xz}^2)/\Delta$ and $V_{0y}\propto(E_{yx}^2 - E_{yz}^2)/\Delta$ contributed from $D_1$ and $D_2$ lines in alkali atoms~\cite{SI}.

\begin{figure}
\includegraphics[scale=0.3]{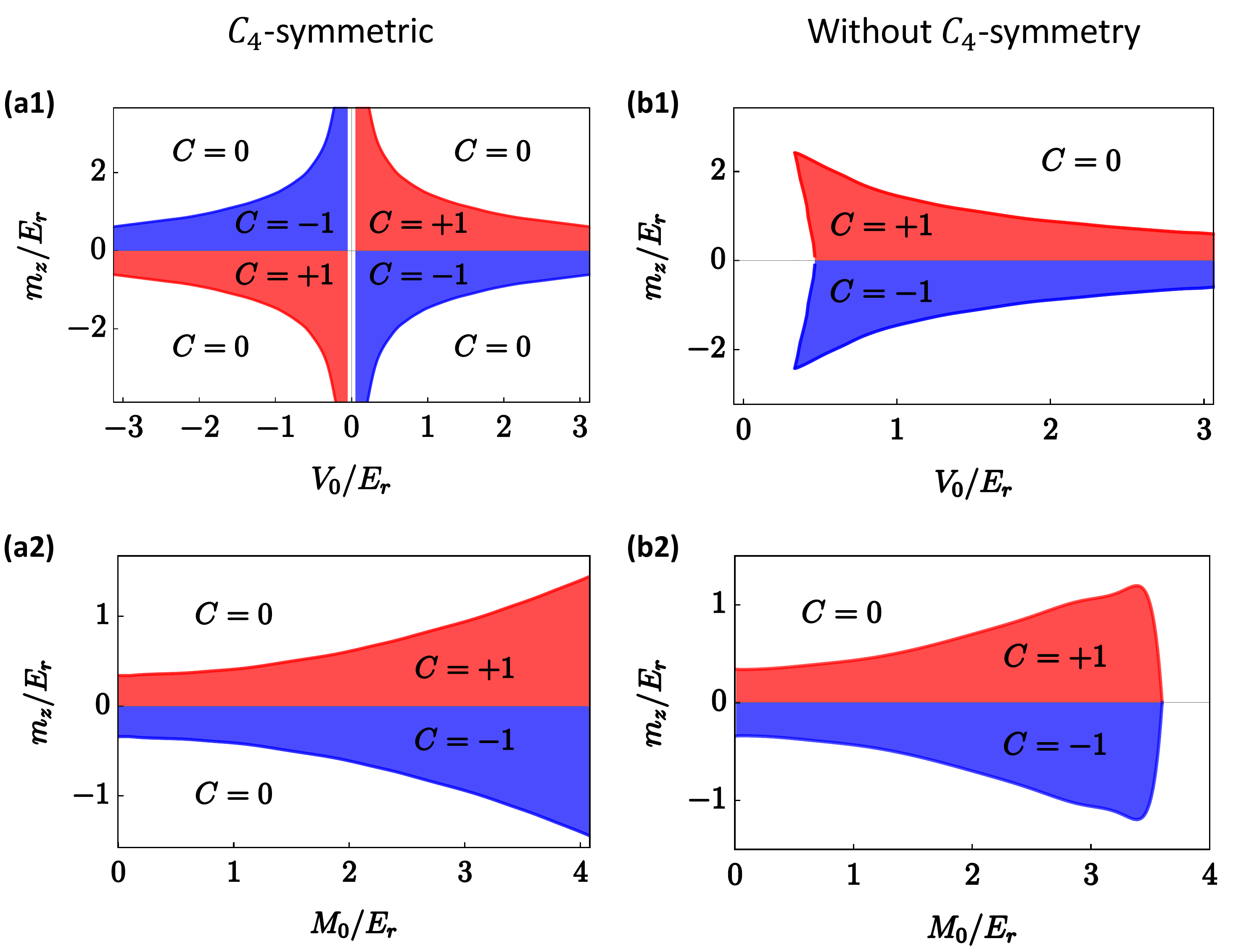}
\caption{Phase diagrams and Chern number for the lowest band,
with the red (blue) regions representing the Chern number $C=+1$ $(-1)$.
The phase diagrams in (a1, b1) are calculated with $M_0 = 1 E_r$ on the $V_0 \text{-} m_z$ plane for $V_0=V_{0x}=V_{0y}$,
and in (a2, b2) are obtained by taking $V_0 = 4 E_r$.}
\label{phase_diag}
\end{figure}

The Raman couplings are also induced by the two lights $\bm{E}_x$ and $\bm{E}_y$, as illustrated in Fig.~\ref{Fig:realization_Dirac}(b), through the $double \text{-} \Lambda$ type configuration.
One Raman potential $\mathcal{M}_x (x,y) = M_{10}\cos k_0y \sin k_0x $, with $M_{10}\propto E_{xz} E_{yx}$, is generated by
the $E_{xy}$ and $E_{yz}$ components,
and the other by $E_{xz}$ and $E_{yx}$ components reads
$\mathcal{M}_y (x,y) = M_{20} \cos k_0x \sin k_0y $, with $M_{20}\propto E_{xy} E_{yz}$. The Raman $\mathcal{M}_x$ ($\mathcal{M}_y$) and lattice potentials satisfy a relative antisymmetric configuration along $x$ ($y$) direction, a crucial property to realize the minimal SO coupled quantum anomalous Hall model~\cite{liu2014realizationtogether,wu2016realization}, which exhibits novel bulk and edge physics~\cite{liu2014topological}. Moreover, the present realization has an exact $C_4$ symmetry: $(x,y;\sigma_x,\sigma_y) \to (y,-x;\sigma_y,-\sigma_x)$, giving $C_4HC_4^{-1}=H$, which is broken by a term $\cos k_0x\cos k_0y(\sigma_{x}+i\sigma_y)$ in the previous scheme~\cite{wu2016realization,SI}. The $C_4$ symmetry has a profound effect on the topological phase boundary, as shown in Fig.~\ref{phase_diag}.
In particular, Fig.~\ref{phase_diag}(a1,a2) and (b1,b2) show the phase diagrams in the $V_0-m_z$ and $M_0-m_z$ planes ($M_0=M_{10}=M_{20}$) for the $C_4$-symmetric and $C_4$-symmetry-breaking systems, respectively. A marked distinction is that the topological region of the $C_4$ symmetric system can expand over the whole $V_0$ and $M_0$ axes, while the $C_4$-symmetry breaking term induces couplings to high orbital bands and reduces the topological region to a finite range versus $V_0$ and $M_0$~\cite{pan2014bose,wu2016realization}.

The key merits of the new scheme are below. First, due to the $C_4$ symmetry, the realization is valid for any type of detuning $\Delta$, blue or red. Moreover, while the phases of two Raman potentials can be random due to the independent light beams along $x$ and $y$ directions, their relative phase is automatically fixed, since ${\cal M}_{x,y}$ are generated from the same two light beams. In contrast, the previous scheme necessitates a long spatial loop to correlate the light beams in $x$ and $y$ directions~\cite{wu2016realization}. These features ensures much greater controllability and stability.

\begin{figure}
\includegraphics[scale=0.255]{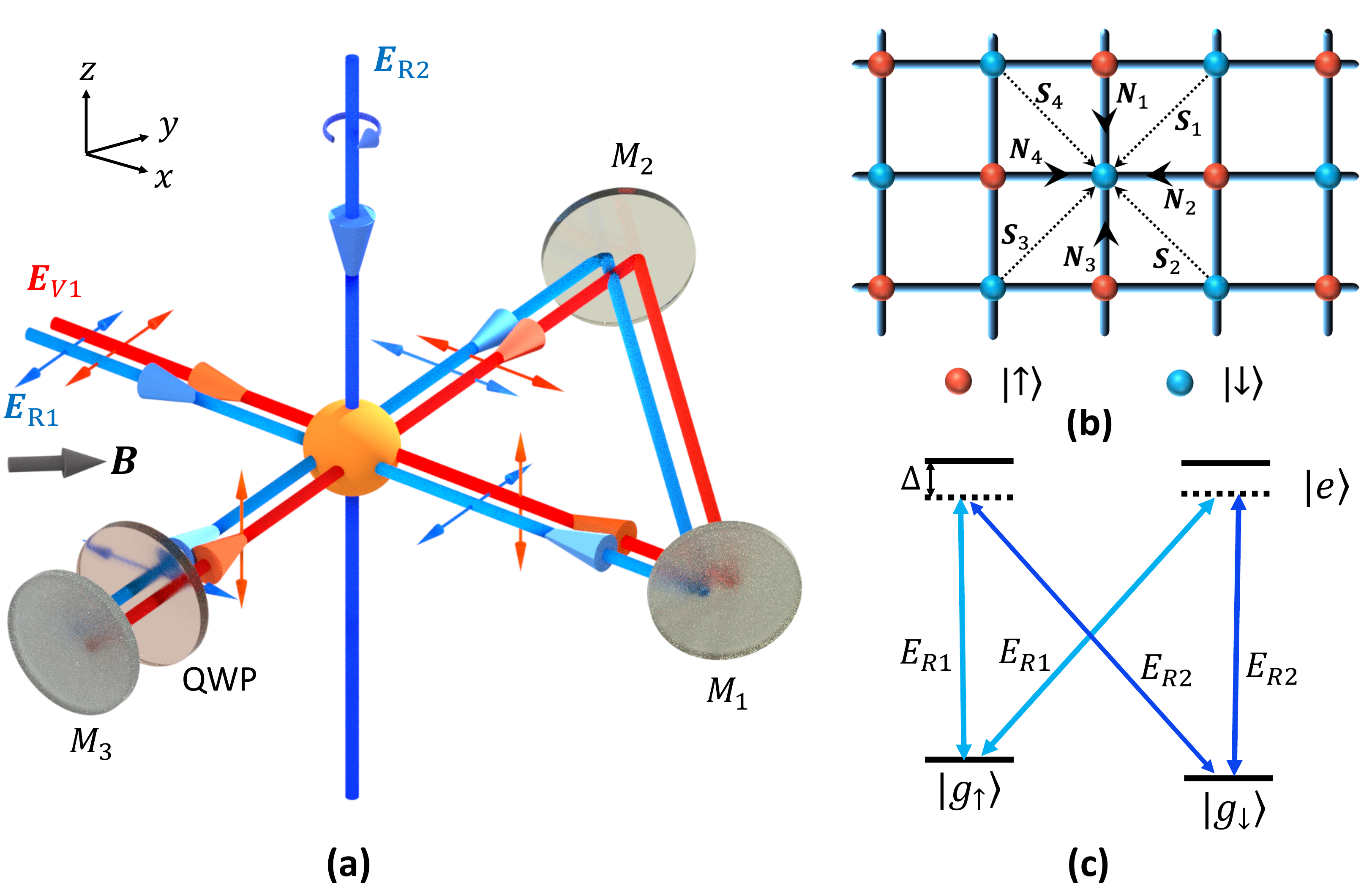}
\caption{(a) Generic setup for realizing Rashba and Weyl type SO couplings. Spin-dependent lattice on $x \text{-} y$ plane are generated by $\bm{E}_{V1}$, and then two independent Raman potentials are generated by $\bm{E}_{R1}$ and $\bm{E}_{R2}$ together. The bias magnetic field is in $x \text{-} y$ plane and has $\pi/4$ angle relative to $x$ axis. (b) Sketch of the tight-binding model of the realization. (c) The scheme of the two independent Raman transitions. }
\label{realization_Rashba}
\end{figure}

\textit{Generalization to Rashba and Weyl SO couplings.}
The above scheme of Dirac type SO coupling can be generalized to Rashba and Weyl types by replacing the spin-independent lattice with spin-dependent one. As showed in Fig.~\ref{realization_Rashba}(a), a laser beam $\bm{E}_{V} (\bm{k}_0,\omega_0)$ is incident along $x$-direction with $y$-polarization, and then reflected by three mirrors. The $\lambda/4$-wave plate before the mirror $M_3$ is applied to rotate the polarization vector of $\bm{E}_{V}$ field by 90 degrees. As a result, the total electric field can be written as $\bm{E}_{V} = \bm{\hat{y}} E_0 e^{ik_0x} + \bm{\hat{z}} E_0 e^{-ik_0x} + \bm{\hat{x}} E_0 e^{-ik_0y} + \bm{\hat{z}} E_0 e^{ik_0y}$. We again neglected the irrelevant phases of the laser beams, which have no effect on the realization~\cite{SI}. The atom-light coupling induces an effective magnetic field $\bm{B}_{\rm eff} \propto \bm{E}^* \times \bm{E} $, for which different hyperfine states experience different lattice potentials $V_{\rm eff}=\bm{B}_{\rm eff} \cdot \bm{F}$~\cite{deutsch1998quantum,mandel2003coherent,mckay2010thermometry}. Thus the lattice potentials of the internal states $|g_{\uparrow} \rangle$ and $|g_{\downarrow} \rangle$ read generically $V_{\sigma}(x,y) = V_{\sigma} \left[ \sin(2k_0x) +  \sin(2k_0y) \right]$, with $\sigma = (\uparrow, \downarrow)$. In experiment, $|g_{\uparrow} \rangle$ and $|g_{\downarrow} \rangle$ are chosen so that $\mathrm{sgn}(V_{\uparrow}) = -\mathrm{sgn}(V_{\downarrow})$, implying that the lattices for $|g_{\uparrow,\downarrow} \rangle$ are staggered [Fig.~\ref{realization_Rashba}(b)~\cite{SI}].

We next show how to generate Raman coupling potentials. As illustrated in Fig.~\ref{realization_Rashba}(a), the first Raman beam $\bm{E}_{R1}(2\bm{k}_0,2\omega_0)$ with double frequency $2\omega_0$ is incident from $x$-direction with $y$-polarization, and running along the same path of  the lattice beam $\bm{E}_{V} $. Being of twice the frequency of $\bm{E}_{V} $, the polarization direction of $\bm{E}_{R1}$ is not affected by $\lambda/4$-wave plate, and therefore it forms a standing wave in $x \text{-} y$ plane: $\bm{E}_{R1} = \bm{\hat{y}} E_{R1}  \cos(2k_0x ) + \bm{\hat{x}} E_{R1} \cos(2k_0y)$. Further, the second Raman beam is incident along $z$-direction $\bm{E}_{R2}(2\bm{k}_0,2\omega_0+\delta \omega) = (i \bm{\hat{x}} + \bm{\hat{y}}) E_{R2} e^{i2k_0z}$. As sketched in Fig.~\ref{realization_Rashba}.(c), $\bm{E}_{R1}$ and $\bm{E}_{R2}$ form the $double \text{-} \Lambda$ type configuration to induce the Raman couplings between $|g_{\uparrow} \rangle$ and $|g_{\downarrow} \rangle$. Similar to the generation of Dirac SO coupling, the Raman potentials take the form $M(x,y,z) = M_1 + M_2 = M_0 e^{i2k_0z}  \left[ \cos(2k_0x) + i \cos(2k_0y) \right]$, where $M_0$ is the amplitude~\cite{SI}. The relative $\pi/2$-phase between $M_1$ and $M_2$ is a consequence of the phase difference between $\bm{\hat{x}}$ and $\bm{\hat{y}}$ components in $\bm{E}_{R2}$, and is crucial to induce the SO coupling in $x \text{-} y$ plane. Moreover, a momentum transfer of $2k_0$ along $z$ direction is induced by $M(x,y,z)$, and leads to additional SO coupling along $z$ direction. Together with a Zemman term $m_z \sigma_z$, the total Hamiltonian is
\begin{align}
H_{\rm 3D} = &{} \frac{\bm{p}^2}{2m} + m_z \sigma_z + V_0 \left( \sin 2k_0x +  \sin 2k_0y \right) \sigma_z  \nonumber  \\
&{} +M_0 e^{i2k_0z\sigma_z}  \left( \cos 2k_0x \sigma_x+ \cos 2k_0y \sigma_y\right).
\end{align}
We show below that $H_{\rm 3D}$ gives rise to 3D Weyl SO coupling. Further, one can reduce the above Hamiltonian to the 2D regime by considering a confinement along $z$ direction. In this case the phase factor $e^{i2k_0z}$ is replaced with a constant, giving a 2D Rashba SO coupling.

The Rashba and Weyl type SO couplings can be best understood with the tight-binding model. We note that the system is in a free space in the $z$ direction, so the tight-binding model shall be derived only in the $x \text{-} y$ plane. Performing a unitary transformation defined by $U_z=e^{-ik_0z\sigma_z}$ yields $H_{\rm 3D}\rightarrow \tilde H_{\rm 3D}=U_zH_{\rm 3D}U^{-1}_z$ that
\begin{align}
\tilde H_{\rm 3D} = &{} \frac{\bm{p}^2}{2m}+ (m_z +\lambda_zp_z)\sigma_z + V_0 \left( \sin 2k_0x +  \sin 2k_0y \right) \sigma_z  \nonumber  \\
&{} +M_0 \left( \cos 2k_0x\sigma_x + \cos 2k_0y\sigma_y \right),
\end{align}
where $\lambda_z=\hbar k_0/m$.
We consider the physics of $s$ orbital $\phi_{s \sigma}(\sigma=\uparrow,\downarrow)$, and take into account the nearest neighbor and diagonal hopping in the $x-y$ plane. The tight-binding Hamiltonian can be straightforwardly written as $H=p_z^2/2m+\lambda_zp_z\sigma_z-t_{\sigma} \sum_{i,j,\sigma} \hat{c}^{\dagger}_{\sigma}(\bm{r_i}) \hat{c}_{\sigma}(\bm{r}_i + \bm{S}_j) + \sum_i m_z ( n_{i \uparrow} - n_{i \downarrow} ) + \sum_{i,j} t_{\rm so}^{i j} \hat{c}^{\dagger}_{\uparrow}(\bm{r}_i) \hat{c}_{\downarrow}(\bm{r}_i + \bm{N}_j) + h.c.$, where $t_{\sigma}$ denotes spin-conserved hopping, $\bm{N}_j$ and $\bm{S_j}$ are vectors connecting a lattice site to its four nearest and four second-nearest neighbor sites, respectively, and $t_{\rm so}^{i j}$ represent spin-flip hopping [Fig.~\ref{realization_Rashba}(b)]. With the basis $\left( \hat{c}_{\bm{q}, \uparrow}, \hat{c}_{\bm{q}, \downarrow} \right)$ the Bloch Hamiltonian reads
\begin{align}
\mathcal{H}_{\rm 3D}(\bm{q}) = &{} \frac{p_z^2}{2m}-4 t_0  \cos(q_x a)  \cos(q_y a)+ (m_z+\lambda_zp_z)\sigma_z  \nonumber  \\
&{} + 2t^{(0)}_{\rm so} {\sigma_x} \sin (q_xa) + 2t^{(0)}_{\rm so} {\sigma_y} \sin (q_ya),
\end{align}
where $a=\pi /(\sqrt{2}k_0)$. For convenience we consider that $t_{\uparrow} = t_{\downarrow} = t_0$ and $a=1$. Expanding the Hamiltonian around the $\Gamma$ point, we reach the generic Weyl type SO coupling $\mathcal{H}(\bm{q})\simeq 4 t_0({q}^2_x + {q}^2_y+\alpha p_z^2) + 2t^{(0)}_{\rm so} ( q_x \sigma_x + q_y \sigma_y+\beta p_z \sigma_z)+m_z\sigma_z$. Here $\alpha=1/(8mt_0)$ and $\beta=\lambda_z/(2t^{(0)}_{\rm so})$. Accordingly, by restricting the system to a 2D plane, Rashba SO coupling can be readily obtained with  $\mathcal{H} (\bm{q})\simeq 2 t_0 (q_x^2 + q_y^2) + t^{(0)}_{\rm so} ( q_x \sigma_x + q_y \sigma_y)+m_z\sigma_z$. For $m_z=0$ the Rashba Hamiltonian respects time-reversal symmetry.

\begin{figure}
\includegraphics[scale=0.33]{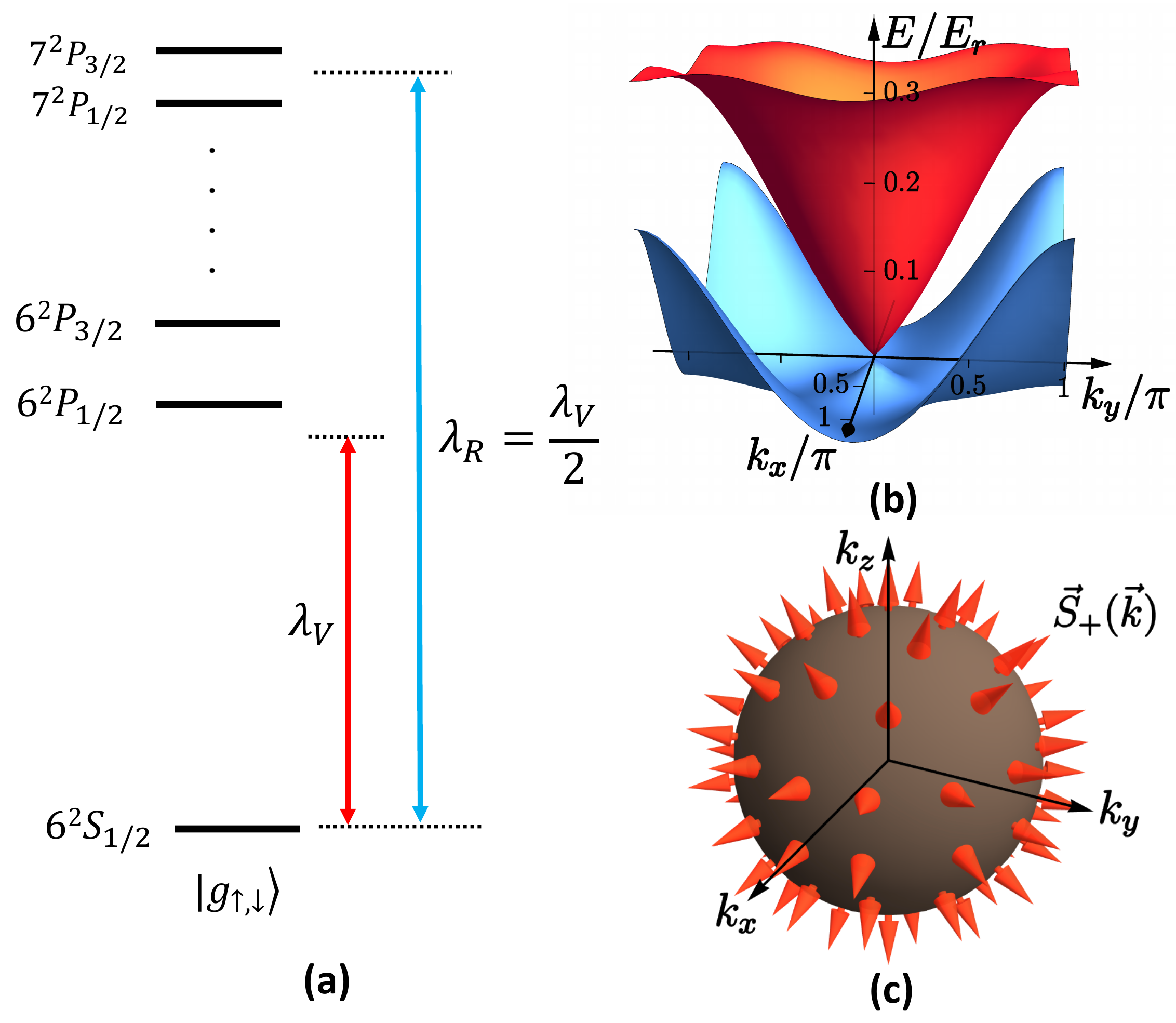}
\caption{(a) Energy levels structure of $^{133}$Cs atoms and the wave length of the lights applied to produce lattice and Raman fields, with $\lambda_V=916.6 nm$ and $\lambda_{R} = 458.3 nm$. (b) The lowest two bands around the $\Gamma$ point for the 2D Rashba SO coupling with $V_0=4E_r$, $M_0=1E_r$ and $m_z=0.68E_r$. (c) Spin texture around the Weyl point $p_z=\bm q=0$.}
\label{Fig4}
\end{figure}

\textit{Realization for $^{133}$Cs Atoms.}
In realizing Rashba and Weyl SO couplings the spin-dependent lattice and Raman couplings are generated by two sources of laser beams with one's frequencies being twice of the other. Note that both spin-dependent lattice and Raman couplings have to be induced by optical transitions with detunings less than the fine structure splitting of excited levels. This is clearly not achievable with a single $D_1/D_2$ transitions. Remarkably, we find surprisingly that $^{133}$Cs atoms provide an ideal candidate for the realization.

Our key idea is to consider the transitions from ground states to both $6^2P_{J}$ and $7^2P_{J}$ levels for generating lattice and Raman potentials [Fig.~\ref{Fig4}(a)]. The ground manifold of ${}^{133}$Cs atoms is $6^2S_{1/2}$, of which we take $|F=4,m_F=-4 \rangle$ and $|3,-3 \rangle$ as spin up $|g_{\uparrow} \rangle$ and spin down $|g_{\downarrow} \rangle$, respectively. The Land\'e $g$-factor $g_F = 1/4$ ($-1/4$) for $|4,-4 \rangle$ ($|3,-3 \rangle$), leading to the opposite lattice potentials by $\bm B_{\rm eff}$ for $|g_{\uparrow} \rangle$ and $|g_{\downarrow} \rangle$.
In particular, the spin-dependent lattice potentials are contributed from transitions to $6^2P_{1/2,3/2}$ states, giving $V(x,y) = (V_0 \bm{\sigma}_z + \delta V_0 \bm{1}) \left( \sin 2k_0x + \sin 2k_0y  \right)$, where $V_0 =\frac{7}{12} E_{V1}^2 \alpha^2_{D_1} /\Delta_1$, with $\alpha_{D_1}$ being the dipole moment of $D_1$ line, and $\delta V_0 \sim 0.1 V_0$ denotes a small distinction of strength of lattice potentials for spin-up and spin-down, with $1/\Delta_1 \equiv 1/\Delta_{1/2} - 1/\Delta_{3/2}$~\cite{SI}. The $\delta V_0$-term corrects $m_z$ in 2D Rashba SO coupling, and shifts the Weyl points in 3D case. The Raman potentials are generated from transitions via $7^2 P_{3/2,1/2}$ states and read $M = M_0 e^{2ik_0z}  \left( \cos 2k_0x + i \cos 2k_0y \right)$, where $M_0 =  \frac{\sqrt{7} \alpha^2_{D_1} }{12 \sqrt{2}} E_{R1} E_{R2}/\Delta_2$, with $\Delta_2$ defined similar as $\Delta_1$~\cite{SI}. We choose the wavelength for Raman ($\lambda_{R}$) and lattice ($\lambda_{V}$) beams $\lambda_{V} = 2 \lambda_{R} = 916.6 nm$. One can verify that the typical parameter condition $(V_0, M_0)= (4E_r, 1E_r)$, with the recoil energy $E_r = \frac{\hbar k_R^2}{2m}$, can be achieved for $^{133}$Cs atoms by taking the lattice beams with a waist $\varpi=200\mu m$ and power $P=15.4mW$, and the Raman beams with the same waist and the power $P=21mW$. This condition can be easily satisfied in experiment.

Fig.~\ref{Fig4}(b) shows the numerical results of the band structure around $\Gamma$ point with the parameters $(4E_r,1E_r)$ and $m_z=0.68E_r$. Interestingly, the band crossing at $\Gamma$ point is obtained, due to the cancellation of $m_z$ by nonzero $\delta V_0$ term at $\Gamma$ point, while the crossings at other symmetric points like $(k_x,k_y)=(\pm\pi,\pm\pi)$ are avoided, implying that the time-reversal symmetry is broken. As detailed in section II.E of supplementary material~\cite{SI}, the $\delta V_0$ term can bring about novel spin texture in momentum space beyond a pure Rashba SO coupling and can have interesting applications. Furthermore, for the 3D Weyl type SO coupling the spin texture around Weyl point is characterized by a skyrmion and protected by Chern numbers, as shown in Fig.~\ref{Fig4}(c)

\textit{Discussion on lifetime and conclusion.} The high controllability and stability are particularly reflected by long lifetime in the present minimized schemes. For the Dirac type SO coupling, we first consider $^{87}$Rb BEC. The heating rate of the dipole trap is about $10nK/s$~\cite{wu2016realization}. The lattice and Raman beams are applied with wavelength $\lambda = 786nm$ (between $D_1$ and $D_2$ lines ) and the heating rate caused by photon scattering is about $63.7nK/s$ for parameters $(4E_r,1E_r)$~\cite{grimm2000optical,SI}. Thus the lifetime of $^{87}$Rb BEC (the time taken for heating to critical temperature) is about $\tau \sim \frac{100nK}{73.7nK/s} \approx 1.36s$. We note that this lifetime has been confirmed in our latest experiment~\cite{experimentalpaper}. For $^{40}$K degenerate fermion gas, we take the lattice and Raman lights with wavelength $\lambda = 768nm$, and can estimate the heating rate to be about $665nK/s$ for parameters $(4E_r,1E_r)$, giving lifetime $\tau \sim 150ms$. The lifetimes are much longer than those in the recent experiments~\cite{wu2016realization,huang2016experimental}.

The lifetime for Rashba and Weyl SO couplings is even much longer for $^{133}$Cs atoms. The condensate lifetime in a typical trap is about $15 s$~\cite{weber2003bose}. Under the parameters $(4E_r,1E_r)$, the heating rate of the lattice beam is only $3nK/s$ due to the large detuning $\Delta_1$, and of Raman beams is also small $\simeq10nK/s$ because of the long lifetime of $7{}^2P_{J}$ manifolds. Thus the lifetime can be $\tau>5s$~\cite{SI}.

In conclusion we have proposed a hierarchy set of minimized experimental schemes to realize Dirac, Rashba, and Weyl SO couplings. These schemes are well accessible with current cold-atom technology, and the resulted SO coupled ultracold atoms can have a long lifetime. For the Dirac type SO coupling, a long-lived Bose-Einstein condensation has been experimentally proved~\cite{experimentalpaper}. For Rashba and 3D Weyl SO couplings, we uncover an ideal candidate provided by $^{133}$Cs atoms for the realization. This work shall essentially push forward cold-atom experimental progresses in realizing high dimensional SO couplings and topological phases in the near future.

The work is supported by the Ministry of Science and
Technology of China (under grants 2016YFA0301604 and 2016YFA0301601), National Natural Science Foundation of China (grant 11574008), Thousand-Young-Talent Program of China, the
CAS, the National Fundamental Research Program (under grant
2013CB922001), and Fundamental Research Funds for the Central
Universities (under grants 2030020028 and 2340000034).

\bibliography{reference}

\onecolumngrid

\renewcommand{\thesection}{S-\arabic{section}}
\setcounter{section}{0}  
\renewcommand{\theequation}{S\arabic{equation}}
\setcounter{equation}{0}  
\renewcommand{\thefigure}{S\arabic{figure}}
\setcounter{figure}{0}  
\newcommand{\diff}{\mathrm{d}}
\newcommand{\biaoti}{\fontsize{12pt}{\baselineskip}\selectfont}

\indent

\section*{\large Supplemental Material}

In this supplementary material we provide the details of deriving the effective Hamiltonians in the main text based on real cold atom candidates, discuss the effects of the different types of spin-orbit (SO) couplings, and calculate the heating and lifetime for the schemes.

\section*{I. Scheme for Dirac Type 2D SO Coupling}

In this section we discuss the details of deriving the effective Hamiltonian. The experimental setup for ${}^{40}$K atoms is shown in Fig. 1 in the main text. Two laser beams form the standing waves
\begin{align}
\bm{E}_{x} &= \bm{\hat{y}} E_{xy} e^{i(\alpha + \alpha_L/2)} \cos(k_0x - \alpha_L/2)  +  \bm{\hat{z}} E_{xz} e^{i(\alpha' + \alpha'_L/2)} \cos(k_0x - \alpha'_L/2)     \nonumber   \\
\bm{E}_{y} &= \bm{\hat{x}} E_{yx}  e^{i(\beta + \beta_L/2)} \cos(k_0y - \beta_L/2) + \bm{\hat{z}} E_{yz}  e^{i(\beta' + \beta'_L/2)} \cos(k_0y - \beta'_L/2).   \nonumber
\end{align}
Here $\alpha, \alpha', \beta, \beta'$ are the initial phases for  $\bm{E}_{xy},\bm{E}_{xz},\bm{E}_{yx},\bm{E}_{yz}$, and $\alpha_L$ is the phase acquired by $\bm{E}_{xy}$ through the optical path from intersecting point to mirror $M_1$, then back to the intersecting point. The phases $\alpha'_L,  \beta_L,  \beta'_L$ have the similar meanings for $\bm{E}_{xz},\bm{E}_{yx},\bm{E}_{yz}$. $E_{xy}, E_{xz}, E_{yx}$, and $E_{yz}$ are real amplitudes. We shall find that the initial phases are irrelevant for the realization.

\subsection{A. Raman Fields}

For convenience, we take ${}^{40}$K atoms for our consideration, while all the results are straightforwardly applicable to $^{87}$Rb atoms. For ${}^{40}$K atoms we define the spin-$1/2$ by the two ground states $|g_{\uparrow} \rangle = |9/2,+9/2\rangle$ and $|g_{\downarrow} \rangle = |9/2,+7/2\rangle$.  Raman coupling scheme for ${}^{40}$K atom is shown in Fig.~\ref{scheme_Dirac}, where two independent Raman transitions are driven by the light components $\bm{E}_{xz},\bm{E}_{yx}$ and $\bm{E}_{xy},\bm{E}_{yz}$. Taking into account the contributions from both $D_1(2^2P_{1/2})$ and $D_2(2^2P_{3/2})$ lines, we have
\begin{align*}
M_1 &= \sum_F \frac{\Omega^{(3/2)*}_{\uparrow F,xz} \cdot \Omega^{(3/2)}_{\downarrow F,yx}}{\Delta_{3/2}} + \sum_F \frac{\Omega^{(1/2)*}_{\uparrow F,xz} \cdot \Omega^{(1/2)}_{\downarrow F,yx}}{\Delta_{1/2}}    \\
&= \frac{\sqrt{2}}{9} \left(  \frac{\alpha^2_{D_1}}{\Delta_{1/2}}  - \frac{\alpha^2_{D_2}}{2 \Delta_{3/2}} \right)  \bm{E}^{*}_{xz} \bm{E}_{yx}^{(+)},
\end{align*}
\begin{align*}
M_2 &= \sum_F \frac{\Omega^{(3/2)*}_{\uparrow F,xy} \cdot \Omega^{(3/2)}_{\downarrow F,yz}}{\Delta_{3/2}} + \sum_F \frac{\Omega^{(1/2)*}_{\uparrow F,xy} \cdot \Omega^{(1/2)}_{\downarrow F,yz}}{\Delta_{1/2}}    \\
&= \frac{\sqrt{2}}{9} \left(  \frac{\alpha^2_{D_1}}{\Delta_{1/2}}  - \frac{\alpha^2_{D_2}}{2 \Delta_{3/2}} \right) \bm{E}^{(-)*}_{xy} \bm{E}_{yz},
\end{align*}
where $\Omega^{J}_{\uparrow F,\gamma z} = \langle \uparrow| ez|F,0,J \rangle \bm{\hat{e}}_z \cdot \bm{E}_{\gamma z}$ and $\Omega^{J}_{\uparrow F,\gamma x} = \langle \uparrow| ex|F,+1,J \rangle \bm{\hat{e}}_+ \cdot \bm{E}_{\gamma x} + \langle \uparrow| ex|F,-1,J \rangle \bm{\hat{e}}_- \cdot \bm{E}_{\gamma x}$, and $\bm{E}_{yx}^{(+)}=\bm{\hat{e}}_+ \cdot \bm{E}_{yx} = {\bm{E}_{yx}}/{\sqrt{2}}$ represents the right-handed light of $\bm{E}_{yx}$, which couples the states $|\frac 92,+\frac 72 \rangle$ and $|F,+\frac 92 \rangle$ as shown in Fig.~\ref{scheme_Dirac}. Similarly, $\bm{E}^{(-)}_{xy}=-i {\bm{E}_{xy}}/{\sqrt{2}}$ represents the left-handed light of $\bm{E}_{xy}$, which couples the states $|\frac 92,+\frac 92 \rangle$ and $|F,+\frac 72 \rangle$. Since $\alpha_{D_2} \approx \sqrt{2} \alpha_{D_1} \approx 5.799 ea_0$~\cite{tiecke2010properties}, with $a_0$ being the Bohr radius, then we have
\begin{align}
M_1 &= M_{10} \cos(k_0x - \alpha'_L/2)  \cos(k_0y - \beta_L/2) e^{-i(\alpha' + \alpha'_L/2)} e^{i(\beta + \beta_L/2)},  \\
M_2 &= i M_{20} \cos(k_0x - \alpha_L/2) \cos(k_0y - \beta'_L/2) e^{-i(\alpha + \alpha_L/2)}  e^{i(\beta' + \beta'_L/2)},
\end{align}
 where
 \begin{align*}
 M_{10/20} = \frac{\alpha^2_{D_1}}{9}\left(\frac{1}{\Delta_{1/2}} - \frac{1}{\Delta_{3/2}}  \right) E_{xz/xy} E_{yx/yz}.
 \end{align*}

To realize a Dirac type 2D SO coupling with nontrivial topology, we require the following two conditions to be satisfied.
(1) The phase difference $\delta \theta$ between $M_1$ and $M_2$ must be non-zero to have a 2D SO coupling. Then the Raman potential can be written as $(|M_1| + |M_2| \cos \delta \theta) \sigma_x + |M_2| \sin \delta \theta \sigma_y $. Here we shall consider the optimal regime with $\delta \theta = \pm \pi/2$, which gives the Raman coupling as $|M_1| \sigma_x \pm |M_2|\sigma_y$ accordingly. Hence the phases should meet the following condition
\begin{align}\label{phase1}
\delta \theta &= (\alpha - \alpha') + (\beta - \beta') + \frac 12 \left[ (\alpha_L - \alpha'_L) + (\beta_L - \beta'_L)  \right] + \pi/2     \nonumber  \\
&= \frac{\pi}{2} + n \pi ,
\end{align}
where $n$ is an integer number. Note that $\alpha$ and $\alpha'$ (similar for $\beta$ and $\beta'$) are phases of two components of the same laser beam. The relative value between them is automatically fixed and can be easily manipulated by wave plates. Also, $\alpha_L$ and $\alpha'_L$ ($\beta_L$ and $\beta'_L$) are the phases acquired through the same optical path, so their relative value is also automatically fixed and can be controlled by wave plates. The system is stable against any phase fluctuations.
(2) To let the SO coupled system to be topological non-trivial, we consider a $\lambda/4$ wave plate before each mirror ($M_1$ and $M_2$). Then the phases will meet the following conditions
\begin{align}\label{phase2}
\frac{1}{2} (\alpha_L - \alpha'_L) = \frac{\pi}{2} + p \pi \ \ {\rm and} \ \ \  \frac{1}{2} (\beta_L - \beta'_L) = \frac{\pi}{2} + q \pi,
\end{align}
with $p$ and $q$ are integer numbers. Accordingly, the Raman potentials turn into
\begin{align}
M_1 = M_{10} \sin(k_0x - \alpha_L/2)  \cos(k_0y - \beta_L/2), \ \ {\rm and} \ \ \ M_2 = \pm i M_{20} \cos(k_0x - \alpha_L/2) \sin(k_0y - \beta_L/2).
\end{align}
We shall see below that the Raman coupling potentials $M_1$ and $M_2$ are antisymmetric with respect to the lattice in $x$ and $y$ directions, respectively, a key point to obtain the nontrivial topology.

In the real experiment the conditions~\eqref{phase1} and~\eqref{phase2} can be satisfied by taking $(\alpha - \alpha') + (\beta - \beta') + \pi/2 = \pi/2 + n \pi$, $(\alpha_L - \alpha'_L)/2 = \pi/2 + p \pi$, and $(\beta_L - \beta'_L)/2 = \pi/2 + q \pi$. Since ${E}_{xy}$ and ${E}_{xz}$ come from the same laser beam, one can naturally take that $\alpha = \alpha'$. Similarly, we have $\beta = \beta'$, and then the phase difference $\delta \theta=\pi/2$. On the other hand, we can control that each $\lambda/4$ wave plate induces an additional $\pi/2$-phase shift to the $\hat z$-component light when the light pass through the wave plate one time. This leads to $(\alpha_L - \alpha'_L)/2=(\beta_L - \beta'_L)/2=\pi/2$, satisfying the condition.

It is trivial to know that the Raman coupling potentials combine to induce a 1D SO coupling when $\delta\theta=0,\pi$. If controlling the phase difference continuously, the crossover between 1D and 2D SO couplings can be induced.

\subsection{B. Spin-independent lattice potentials}

\begin{figure}
\includegraphics[scale=0.33]{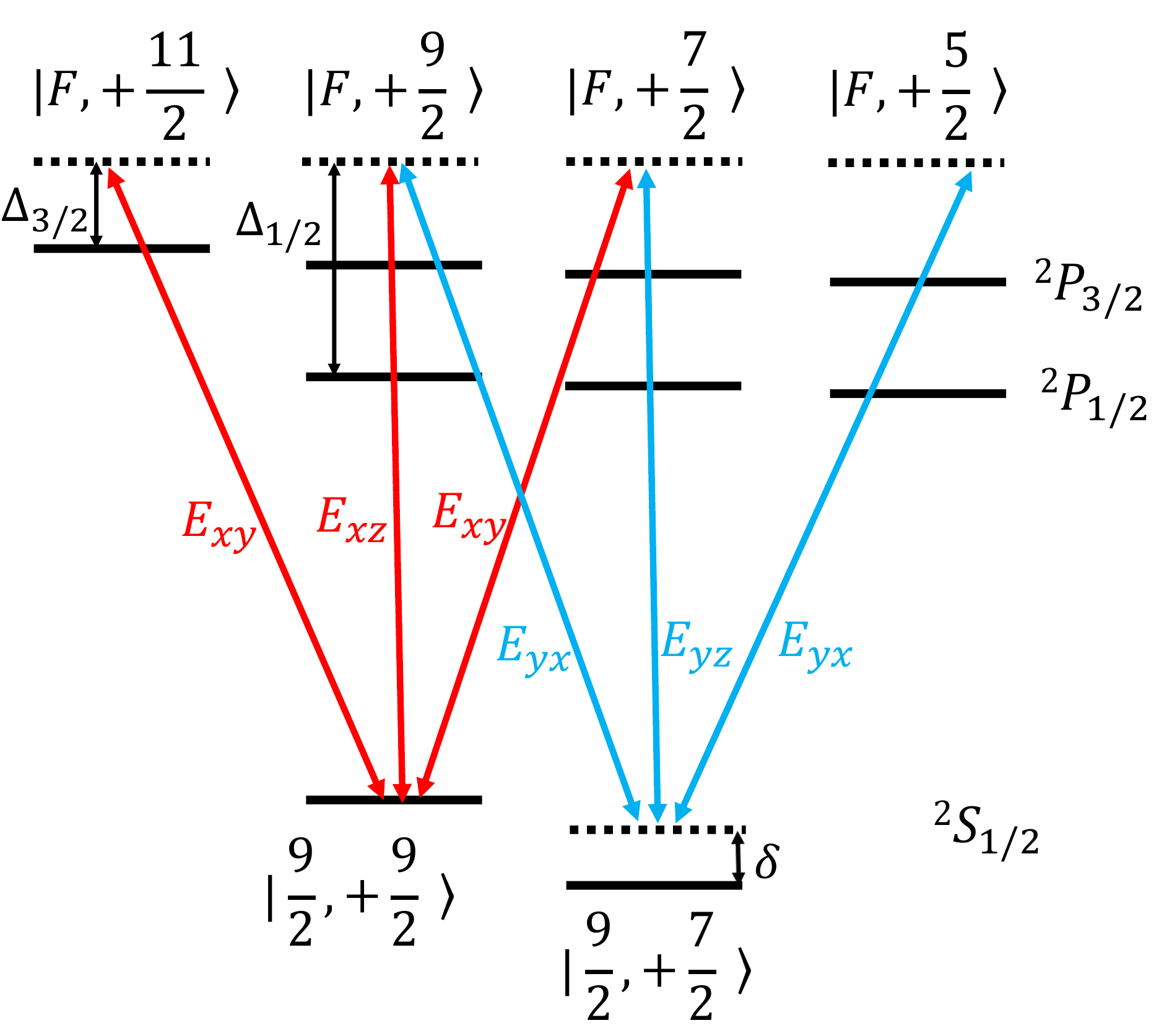}
\caption{Spin-independent lattice potential induced by four linearly polarized standing wave lights which couple the $D_1$ and $D_2$ transitions.}
\label{scheme_Dirac}
\end{figure}

As shown in Fig.~\ref{scheme_Dirac}, the optical lattice is contributed from both $D_2$ $(2^2 S_{1/2} \to 2^2 P_{3/2})$ and $D_1$ $(2^2 S_{1/2} \to 2^2 P_{1/2})$ lines, and it is a spin-independent lattice
\begin{align}
V_{\uparrow} = V_{\downarrow} &= \sum_F \frac{1}{\Delta_{3/2}} \left( |\Omega^{(3/2)}_{\uparrow F,xz}|^2 +  |\Omega^{(3/2)}_{\uparrow F,xy}|^2 +  |\Omega^{(3/2)}_{\uparrow F,yz}|^2 +  |\Omega^{(3/2)}_{\uparrow F,yx}|^2  \right)    \nonumber  \\
&{} \quad + \sum_F \frac{1}{\Delta_{1/2}} \left( |\Omega^{(1/2)}_{\uparrow F,xz}|^2 +  |\Omega^{(1/2)}_{\uparrow F,xy}|^2 +  |\Omega^{(1/2)}_{\uparrow F,yz}|^2 +  |\Omega^{(1/2)}_{\uparrow F,yx}|^2  \right)    \nonumber  \\
&= V_{0x} \cos^2(k_0x - \alpha_L/2) + V_{0y} \cos^2(k_0y - \beta_L/2) ).
\end{align}
Here the constant is neglected in the last line, and we have $V_{0x} = \frac{\alpha_{D_1}^2}{3} (\frac{2}{\Delta_{3/2}} + \frac{1}{\Delta_{1/2}})( E_{xy}^2 - E_{xz}^2 )$ and $V_{0y}= \frac{\alpha_{D_1}^2}{3} (\frac{2}{\Delta_{3/2}} + \frac{1}{\Delta_{1/2}}) (E_{yx}^2 - E_{yz}^2)$. Note that when $E_{xy} > E_{xz}$ and $E_{yx} > E_{yz}$, the lattice potential takes the form $\cos^2(k_0x) + \cos^2(k_0y)$. In comparison, when $E_{xy}<E_{xz}$ and $E_{yx}<E_{yz}$, the lattice potential becomes $\sin^2(k_0x) + \sin^2(k_0y)$, which corresponds to a translation along the diagonal direction compared to $\cos^2(k_0x) + \cos^2(k_0y)$. Effectively, the sign of the optical potentials is reversed for the two cases, equivalent to changing the sign of the detuning (e.g. from blue- to red-detuned or vise versa).

\subsection{C. Effective Hamiltonian}

As clarified previously, the phase fluctuations for lattice and Raman fields are the same: $\phi^{\rm fluc}_x = - \alpha_L/2$, and $\phi^{\rm fluc}_y = - \beta_L/2$. Hence the relative spatial configuration of $M_{x/y}$ and $V$ are always automatically fixed, and the fluctuations  only lead to a global shift of the lattice and Raman fields. Therefore, we can set $\alpha_L =  \beta_L = 0$ safely. The global phase of Raman potentials are irrelevant and can also be removed. Then the Hamiltonian can be written as
\begin{align}
H=\frac{p^2}{2m} + V_{0x} \cos^2(k_0x) + V_{0y} \cos^2(k_0y) + \mathcal{M}_x \sigma_x + \mathcal{M}_y \sigma_y + m_z \sigma_z,
\end{align}
the strength of Raman coupling $\mathcal{M}_x = |M_1| + |M_2| \cos \delta \theta$ and $\mathcal{M}_y = |M_2| \sin \delta \theta$, which reduce to an 1D SO coupling for $\delta \theta = n \pi$ and an optimal 2D Dirac type SO coupling for $\delta \theta = \pi/2 + n \pi$. This enables a fully controllable study of the crossover between 2D and 1D SO couplings by tuning $\delta \theta$.

\section*{II. Scheme for Rashba and Weyl Type SO Couplings}

\subsection{A. Spin-dependent lattice Potential}

\begin{figure}
\centering \includegraphics[scale=0.35]{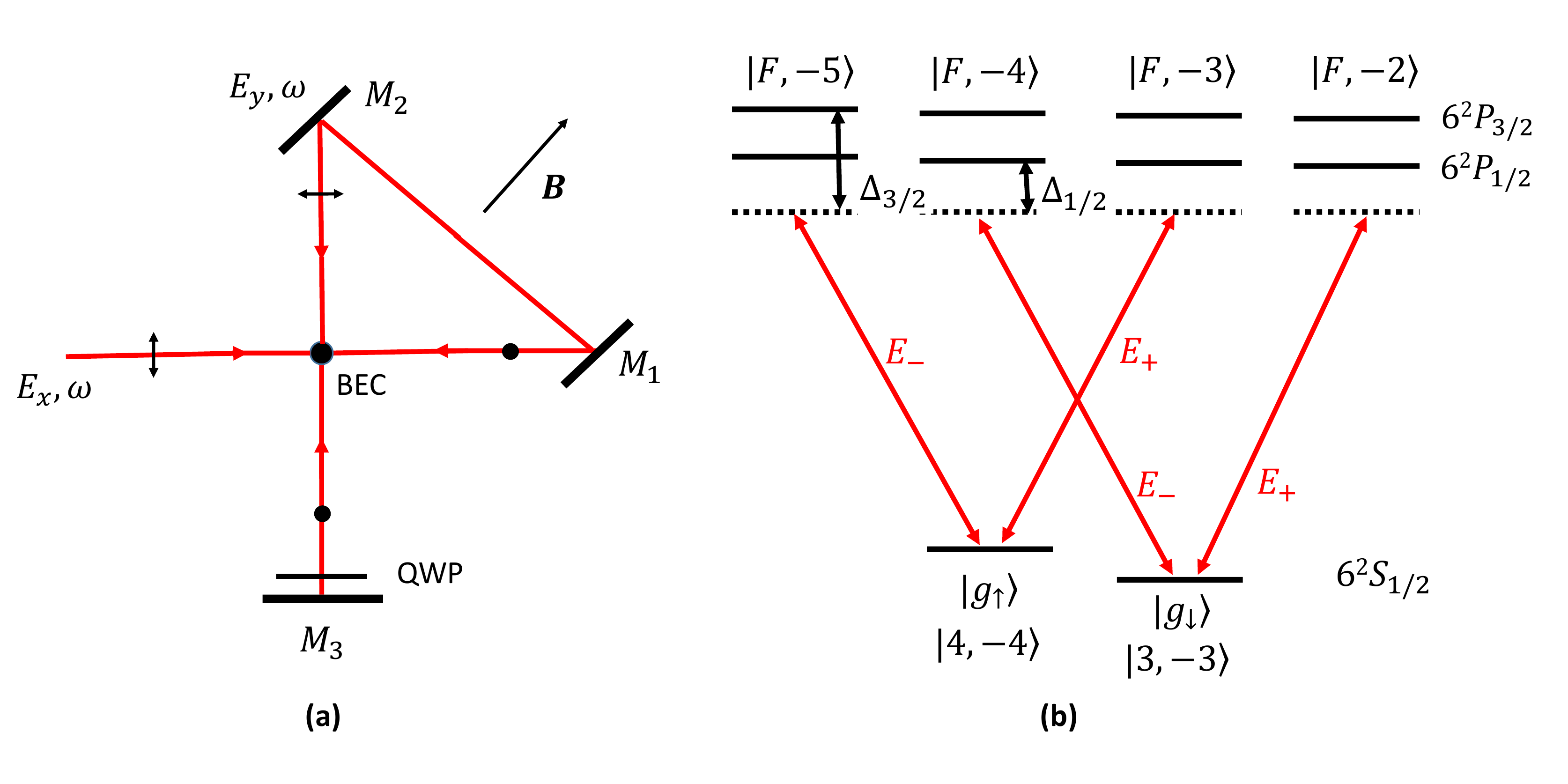}
\caption{(a) Experiment setting for spin-dependent lattice potential. The lattice potential in each ($x$ and $y$) direction is formed by two lights traveling in opposite directions with mutual perpendicular polarization. (b) Optical transitions allowed by selection rule for generating spin-dependent lattice potential along $x/y$ direction.}
\label{lattice_Rashba}
\end{figure}

We generate the spin-dependent square lattice potential in $x \text{-} y$ plane from the traveling-wave beams, described by the electric field $\bm{E}_{V}$
\begin{align*}
\bm{E}_{Vx} &= \bm{\hat{y}} E_0 e^{ik_0x+i\phi_0} + \bm{\hat{z}} E_0 e^{-ik_0x + i\phi_0 + 2i\phi_L + i\phi'_L}     \\
\bm{E}_{Vy} &= \bm{\hat{x}} E_0 e^{-ik_0y + i\phi_0 + i\phi_L} + \bm{\hat{z}} E_0 e^{ik_0y + i\phi_0 + i\phi_L + i\phi'_L},
\end{align*}
where $\bm{E}_{V,x/y}$ represents the laser propagating along $x/y$-direction, $\phi_0$ is the initial phase of the laser beam, and $\phi_L/\phi'_L$ is the phase acquired on path $L/L'$. The path $L$ denotes the loop from lattice center to mirror $M_1$, then to $M_2$, and finally back to the lattice center, while the path $L'$ denotes the one from lattice center to the mirror $M_3$ and back to lattice center [Fig.~\ref{lattice_Rashba}(a)].

In general, the optical potential generated for atoms in the ground state is related to the electromagnetic field by
\begin{align}
V(r) = u_s |\bm{E}|^2 + i u_v \left( \bm{E}^* \times \bm{E}  \right) \cdot \bm{S},
\end{align}
where the first term is the scalar potential with $u_s=-\frac{1}{12 \Delta_s} | \langle l=0 | \bm{d} | l=1 \rangle |^2 $ and $ u_s |\bm{E}|^2$, and the second term denotes a vector light shift, with $u_v = \frac{A_{FS}}{\hbar \Delta_e}  u_s = \frac{2 \Delta_{FS}}{3 \Delta_e} u_s $ and $\Delta_{FS}$ being the fine-structure splitting. The scalar potential is spin-independent for linearly polarized lights. In comparison, the vector light shift is spin-dependent. The effective Hamiltonian $H_{\rm eff} = u_s |\bm{E}|^2 + \frac{\mu_B g_J}{\hbar} \left( \bm{B} + \bm{B}_{\rm eff}  \right) \cdot \bm{J}$, where we have replaced $\bm{S}$ with $\bm{J} = \bm{S} + \bm{L}$ since  $\bm{L}=0$ for the ground state. The symbol $\bm{B}$ denotes the external magnet field, and  $\bm{B}_{\rm eff} = \frac{i \hbar u_v \left( \bm{E}^* \times \bm{E}  \right) }{\mu_B g_J}$ is the effective magnetic field. Considering the hyperfine structure, we further replace $g_J \bm{J}$ with $g_F \bm{F}$ and then have
\begin{align}
H_{\rm eff} = u_s |\bm{E}|^2 + \frac{\mu_B g_F}{\hbar} \left( \bm{B} + \bm{B}_{\rm eff}  \right) \cdot \bm{F}.
\end{align}
In our proposal, the effective magenetic field produced by the lasers has the equal components in $\bm{x}$ and $\bm{y}$ directions. Therefore, we must apply external magnetic field along the diagonal direction ($B_x^{\rm eff},B_y^{\rm eff}\neq0$) of $x-y$ plane. This ensures that $\bm{B}_{\rm eff}$ has nonzero components along the direction of $\bm{B}$.

With this setup, the effective magnetic field should be $B_x^{\rm eff}\hat e_x + B_y^{\rm eff}\hat e_y$. The external magnetic field $\bm{B}$ along the diagonal direction satisfies $B \gg B_x^{\rm eff}, B_y^{\rm eff}$. Thus, the total field reads $B_t =\sqrt{(B + \frac{B_x^{\rm eff}}{\sqrt{2}} + \frac{B_y^{\rm eff}}{\sqrt{2}} )^2 + (\frac{B_x^{\rm eff}}{\sqrt{2}} - \frac{B_y^{\rm eff}}{\sqrt{2}})^2}$. We expand the total field $B_t$ around the point $B_x^{\rm eff} = B_y^{\rm eff}=0$, keeping the first-order term, and finally have $B_t \approx B +\frac{B_x^{\rm eff}}{\sqrt{2}} +  \frac{B_y^{\rm eff}}{\sqrt{2}} $. In other words, if we decompose the effective magnetic field in the traverse and longitudinal direction of the external magnetic field, only the longitudinal component matters.

Next we focus on the expression of the effective magnetic field
\begin{align*}
\bm{B}_{\rm eff}  &\propto     \bm{E}^* \times \bm{E} \\
& = ( \bm{E}^*_{Vx} \times \bm{E}_{Vx} + \bm{E}^*_{Vy} \times \bm{E}_{Vy} ) +  ( \bm{E}^*_{Vx} \times \bm{E}_{Vy} + \bm{E}^*_{Vy} \times \bm{E}_{Vx} ) \\
& =( \bm{B}_{x}^{\rm eff} +  \bm{B}_{y}^{\rm eff} )+ \bm{B}_{xy}^{\rm eff}.
\end{align*}
The cross term $\bm{B}_{xy}^{\rm eff}$ can be obtained
\begin{align*}
\bm{B}_{xy}^{\rm eff} \propto \bm{\hat{y}} E_0^2 \sin(k_0x-k_0y-\phi_L-\phi'_L) -\bm{\hat{x}} E_0^2 \sin(k_0x-k_0y-\phi_L-\phi'_L) + \bm{\hat{z}}E_0^2 \sin(k_0x+k_0y-\phi_L).
\end{align*}
However, this term is perpendicular to external field $\bm{B}$, so the projection of this term in the direction of $\bm{B}$ is zero, hence it can be neglected. It is ready to verify that the scalar potential is a constant, and can be neglected. Finally the lattice potential takes the form qualitatively
\begin{align}
V_{\rm latt} = ( \bm{B}_{x}^{\rm eff} +  \bm{B}_{y}^{\rm eff} ) \cdot \bm{F}  \propto  E^2_0 \sin(2k_0x - 2\phi_L - \phi'_L) + E^2_0 \sin(2k_0y + \phi'_L).
\end{align}

Note that we can use another method to obtain the result more accurately. We first consider the laser propagating along $x$-direction, write the electric fields in the basis of the circular polarized light
\begin{align*}
E_{Vx} &= \frac{E_0}{\sqrt{2}} e^{i \phi_0 + i\phi_L + i\phi'_L/2 +i\pi/4} 2\left[ i \bm{\hat{e}}_{+} \sin(k_0x - \phi_L - \phi'_L/2 - \pi/4) + \bm{\hat{e}}_-  \cos(k_0x - \phi_L - \phi'_L/2 - \pi/4) \right]    \nonumber   \nonumber  \\
&= \bm{\hat{e}}_+ E_{+} +  \bm{\hat{e}}_- E_{-},
\end{align*}
and
\begin{align*}
|E_{+}|^2 &= 2 E^2_0 \sin^2(k_0x - \phi_L -\phi'_L/2 - \pi/4) = E^2_0 - E^2_0 \sin(2k_0x - 2\phi_L -\phi'_L)   \nonumber  \\
|E_{-}|^2 &= 2 E^2_0 \cos^2(k_0x - \phi_L -\phi'_L/2 - \pi/4) = E^2_0 + E^2_0 \sin(2k_0x - 2\phi_L -\phi'_L) .
\end{align*}

As shown in Fig.~\ref{lattice_Rashba}, the spin-dependent lattice potential is contributed from both $D_2$ $(6^2 S_{1/2} \to 6^2 P_{3/2})$ and $D_1$ $(6^2 S_{1/2} \to 6^2 P_{1/2})$ lines~\cite{steck2003cesium}. The light $\bm{E}_{V}$ drive the $\sigma$ transitions from ground states $|F=4,m_F=-4 \rangle$ and $|F=3,m_F=-3 \rangle$ to all possible excited levels which satisfy the selection rule. The state $|4,-4 \rangle$ $(|g_{\uparrow} \rangle)$ is coupled to excited states $|F,-5 \rangle$ and $|F,-3 \rangle$, while the state $|3,-3 \rangle$ $(|g_{\downarrow} \rangle)$ is coupled to other states $|F,-4 \rangle$ and $|F,-2 \rangle$. The detunings $\Delta_{1/2}$ and $\Delta_{3/2}$ are much larger than the hyperfine-structure splitting, and have the same order of magnititude of the fine-structure splitting, which is the  energy difference between $D_1$ and $D_2$ lines. Then the lattice potential can be obtained by summing over the contributions of all the allowed transitions
\begin{align}
V_{\uparrow}(x) &= \sum_F \frac{1}{\Delta_{3/2}} \left( |\Omega^{(3/2)}_{\uparrow F,+1}|^2 + |\Omega^{(3/2)}_{\uparrow F,-1}|^2  \right) + \sum_F \frac{1}{\Delta_{1/2}} \left( |\Omega^{(1/2)}_{\uparrow F,+1}|^2 + |\Omega^{(1/2)}_{\uparrow F,-1}|^2  \right)   \nonumber  \\
&=  \frac{2}{3} E_{\rm V}^2  \alpha^2_{D_1}  \left( \frac{1}{\Delta_{3/2}} - \frac{1}{\Delta_{1/2}}  \right) \sin(2k_0x - 2\phi_L -\phi'_L).
\end{align}
We have neglected all the constants in the final result. Taking $V_0 = E_{0}^2  \alpha^2_{D_1}/\Delta_1$, with $1/\Delta_1= \left( 1/\Delta_{3/2} - 1/\Delta_{1/2}  \right)$, we have $V_{\downarrow}(x) = - \frac{1}{2} V_0  \sin(2k_0x - 2\phi_L -\phi'_L)$ for the atoms staying in the state $|3,-3 \rangle$. Similarly,  the spin-dependent lattice potential in $y$-direction is $V_{\uparrow}(y) = \frac{2}{3} V_0  \sin(2k_0y + \phi'_L)$ and $V_{\downarrow}(y) = - \frac{1}{2} V_0 \sin(2k_0y + \phi'_L)$. The total lattice can be finally written as
\begin{eqnarray}
V_{\rm latt} =  \left( \frac{1}{12} \bm{1} + \frac{7}{12} \bm{\sigma_z} \right) V_0  \left[  \sin(2k_0x - 2\phi_L -\phi'_L) +  \sin(2k_0y + \phi'_L)  \right].
\end{eqnarray}

\subsection{B. Raman Fields}
\begin{figure}
\includegraphics[scale=0.35]{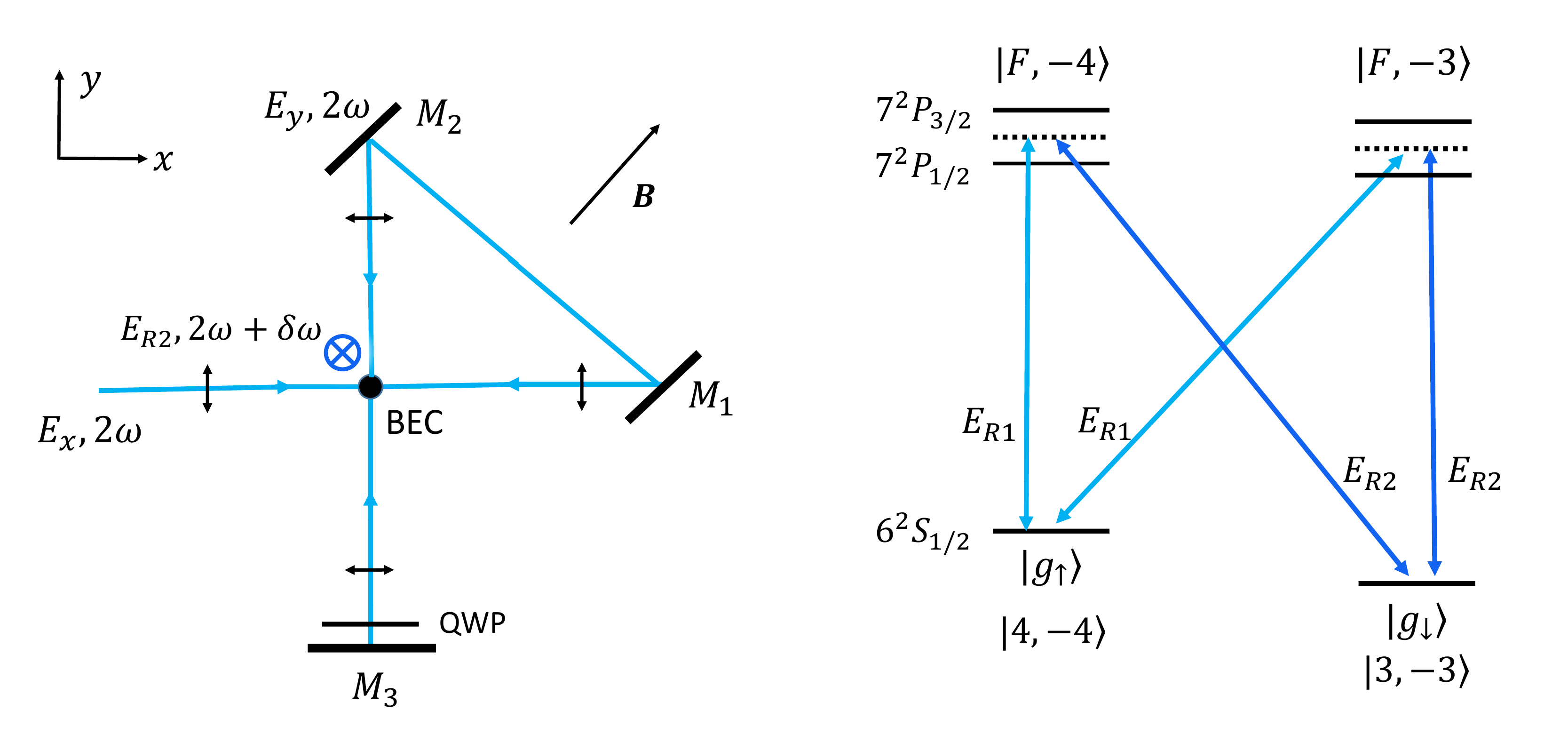}
\caption{Experiment setting and coupling scheme for Raman couplings.}
\label{Raman_Rashba}
\end{figure}

Now we study the generation of Raman fields by adding another two frequency-doubled lights $\bm{E}_{R1}(2\bm{k}_0,2\omega_0)$ and $\bm{E}_{R2}(2\bm{k}_0,2\omega_0+\delta \omega)$, where $\delta \omega$ is of the same order of the energy difference between $|g_{\uparrow} \rangle$ and $|g_{\downarrow} \rangle$. The polarization of $\bm{E}_{R1}$ can't be affected by $\lambda/4$ wave plate, hence $\bm{E}_{R1}$ form a standing wave on $x$-$y$ plane, and  $\bm{E}_{R2}$ is just a travelling wave, i.e.
\begin{align*}
\bm{E}_{R1} &= E_{R1} e^{i\phi_1 + 2i\phi_L + i\phi'_L} \left[ \bm{\hat{y}} \cos(2k_0x - 2\phi_L - \phi'_L) + \bm{\hat{x}} \cos(2k_0y + \phi'_L) \right],   \nonumber  \\
\bm{E}_{R2} &= ( i \bm{\hat{x}} +   \bm{\hat{y}} ) E_{R2} e^{2ik_0z + i\phi_2}.
\end{align*}
The Raman fields are contributed from $6^2 S_{1/2} \to 7^2 P_{3/2}$ and $6^2 S_{1/2} \to 7^2 P_{1/2}$ transitions. It's a little tedious to calculate Raman potentials because the magnetic field points to diagonal direction. We first need to decompose the vectors $\bm{E}_{R1}, \bm{E}_{R2}$ along the external magnetic field direction
\begin{align*}
\bm{E}_{R1} &= \frac{1}{\sqrt{2}} E_{R1} e^{i \theta} \left[ \cos(2k_0x - 2\phi_L - \phi'_L) + \cos(2k_0y + \phi'_L)  \right] \bm{\hat{B}}  + \frac{1}{\sqrt{2}} E_{R1} e^{i \theta} \left[ \cos(2k_0x - 2\phi_L - \phi'_L) - \cos(2k_0y + \phi'_L)  \right]  \bm{\hat{B}}_{\perp},   \nonumber \\
\bm{E}_{R2} &= \frac{E_{R2}}{\sqrt{2}} \left( 1 + i \right) e^{2ik_0z + i\phi_0} \bm{\hat{B}} + \frac{E_{R2}}{\sqrt{2}} \left( 1-i \right) e^{2ik_0z + i\phi_0} \bm{\hat{B}}_{\perp},
\end{align*}
where $\theta = \phi_0 + 2 \phi_L + \phi'_L$. Then the potentials can be obtained as follows
\begin{align}
M_1 &= \sum_{F} \frac{\Omega^{(3/2)*}_{\uparrow F,1\sslash} \cdot \Omega^{(3/2)}_{\downarrow F,2\perp} }{\tilde \Delta_{3/2}} + \sum_F \frac{ \Omega^{(1/2)*}_{\uparrow F,1\sslash} \cdot \Omega^{(1/2)}_{\downarrow F,2\perp} }{\tilde \Delta_{1/2}}   \nonumber  \\
&= \frac{\sqrt{7}  \alpha^2_{\tilde D_1} }{12 \sqrt{2}} \left( \frac{1}{\tilde \Delta_{1/2}} - \frac{1}{ \tilde \Delta_{3/2}}  \right) E_{R1} E_{R2} (1-i) e^{2ik_0z} [ \cos(2k_0x - 2\phi_L - \phi'_L) + \cos(2k_0y + \phi'_L)],
\end{align}
and
\begin{align}
M_2 &= \frac{\sqrt{7}  \alpha^2_{D_1} }{12 \sqrt{2}} \left( \frac{1}{\tilde \Delta_{1/2}} - \frac{1}{\tilde  \Delta_{3/2}}  \right) E_{R1} E_{R2} (1+i) e^{2ik_0z} \left[ \cos(2k_0x - 2\phi_L - \phi'_L) - \cos(2k_0y + \phi'_L) \right].
\end{align}
Note that in general the detunings $\tilde \Delta_{1/2,3/2}\neq\Delta_{1/2,3/2}$. As discussed in the main text, lattice potential $V$ and Raman fields $M$ have the same phase fluctuations, so the fluctuations can be neglected and for simplicity we take $\phi^{fluc} = 0$. We then find
\begin{align}
M = M_1 + M_2 &= M_0 e^{2ik_0z}  \left( \cos 2k_0x + i \cos 2k_0y \right),
\end{align}
where $M_0 =\frac{\sqrt{7}  \alpha^2_{D_1} }{6 \sqrt{2}} E_{R1} E_{R2}/\Delta_2$, with $\frac{1}{\Delta_2}=\left( \frac{1}{\Delta_{1/2}} - \frac{1}{ \Delta_{3/2}}  \right) $.

\subsection{C. Effective Hamiltonian}
With the lattice and Raman potentials above, the total Hamiltonian can be written as 
\begin{align}
H_{\rm 3D} = &{} \frac{\bm{p}^2}{2m}  + m_z \sigma_z  \nonumber + V_0 \left( \sin 2k_0x +  \sin 2k_0y \right) \sigma_z + M_0 \left[ e^{i2k_0z} ( \cos 2k_0x + i \cos 2k_0y )|g_{\uparrow}\rangle \langle g_{\downarrow}| +h.c. \right].
\end{align}
To remove the phase term $e^{i2k_0z}$ in the Raman potentials in order to make the Hamiltonian periodic, we define a rotation operator $U=e^{-ik_0z \sigma_z}$ and perform a transformation
\begin{align*}
H_{\rm 3D} \to \tilde{H}_{\rm 3D} = U H_{\rm 3D} U^{\dagger}.
\end{align*}
Then the Hamiltonian would change into
\begin{align}
\tilde{H}_{\rm 3D} = &{} \frac{\bm{p}^2}{2m}  + (m_z + \lambda_z p_z) \sigma_z  \nonumber + V_0 \left( \sin 2k_0x +  \sin 2k_0y \right) \sigma_z + M_0 \left( \cos 2k_0x \sigma_x + \cos 2k_0y \sigma_y \right) ,
\end{align}
where $\lambda_z = \hbar k_0/m$ and we have ignored the constant above.

\subsection{D. The Tight-Binding Model}
\begin{figure}
\centering \includegraphics[scale=0.35]{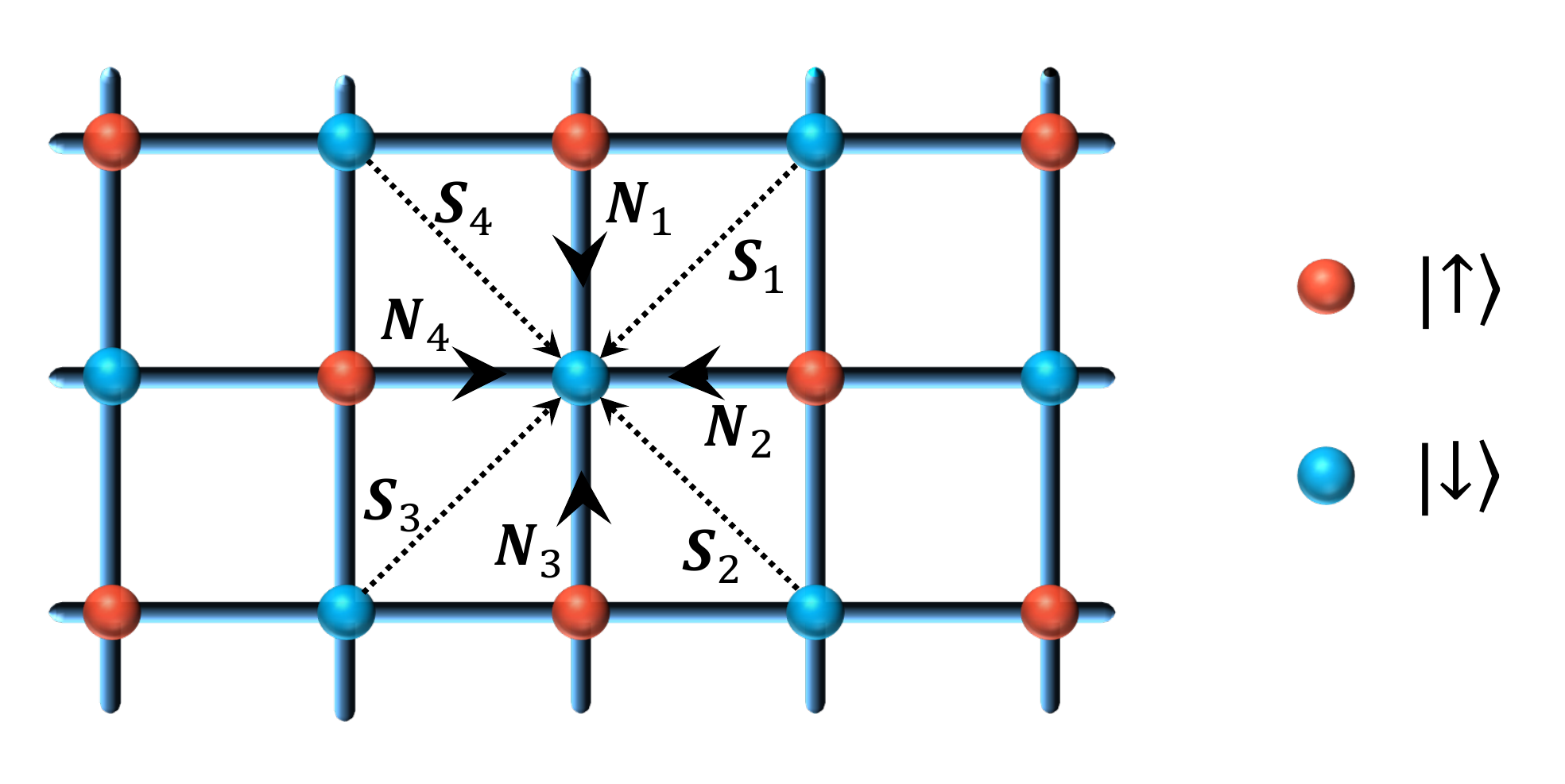}
\caption{Sketch for the tight-binding model, with atoms trapped in a spin-dependent square optical lattice. We only consider the nearest and second-nearest (diagonal) neighbor hopping couplings.}
\label{TBM}
\end{figure}

We next derive the tight-binding model from the effective Hamiltonian, while we emphasize that the realization of SO couplings is not restricted by tight-binding model. For simplicity, we first discuss the tight-binding model in 2D system, and then extend it to 3D. We only consider the nearest and next-nearest neighbor hopping, and assume that atoms stay in the lowest $s$ orbits. As shown in Fig.~\ref{TBM}, the tight-binding model of this system in real space can be written as
\begin{align*}
H=-t_{\sigma} \sum_{i,j,\sigma} \hat{c}^{\dagger}_{\sigma}(\bm{r}_i) \hat{c}_{\sigma}(\bm{r}_i + \bm{S}_j) + \sum_i m_z ( \hat{n}_{i \uparrow} - \hat{n}_{i \downarrow} ) + \sum_{i,j} t_{\rm so}^{{i} {j}} \hat{c}^{\dagger}_{\uparrow}(\bm{r}_i) \hat{c}_{\downarrow}(\bm{r}_i + \bm{N}_j) + h.c.
\end{align*}
where $\bm{N}_i$ and $\bm{S}_i$ represents the distances between the nearest neighbor sites and second-nearest neighbor sites, respectively. The particle number operators are defined as $\hat{n}_{i \sigma} = \hat{c}^{\dagger}_{i,\sigma} \hat{c}_{i,\sigma}$, and $t_{\sigma} (\sigma=\uparrow,\downarrow) $ denotes the spin-conserved hopping, given by
\begin{align*}
t_{\sigma} = \int \diff^2 \bm{r} \phi_{s\sigma}^{(i)} (\bm{r}) \left[ \frac{\bm{p}^2}{2m} + V(\bm{r}) \right] \phi_{s\sigma}^{(j)}(\bm{r}).
\end{align*}
 $t_{\rm so}^{ij}$ is the spin-flipped hopping coefficient given by
 \begin{align*}
 t_{\rm so}^{i j} =  \int \diff^2 \bm{r} \phi_{s \uparrow}^{(i)} (\bm{r}) M_{1/2}(\bm{r}) \phi_{s \downarrow}^{(j)}(\bm{r}),
\end{align*}
 representing the induced 2D SO interaction. It can be readily verified that the spin-flip hopping terms due to Raman fields satisfy $t_{\rm so}^{j_x,j_x \pm 1} = \pm t_{\rm so}^{(0)}$ and $t_{\rm so}^{j_y,j_y \pm 1} = \pm i t_{\rm so}^{(0)}$. Then we transform the above equation into momentum space and write the  Hamiltonian in matrix form
\begin{eqnarray*}
H &=& \sum_{\bm{q}}
\left(
\begin{array}{cc}
\hat{c}^{\dagger}_{\bm{q}, \uparrow} & \hat{c}^{\dagger}_{\bm{q}, \downarrow}
\end{array}
\right)
\mathcal{H}(\bm{q})
\left(
\begin{array}{c}
\hat{c}_{\bm{q}, \uparrow}  \\
\hat{c}_{\bm{q}, \downarrow}
\end{array}
\right).
\end{eqnarray*}
By setting $t_{\uparrow} = t_{\downarrow} = t_0$, the matrix $\mathcal{H}$ in momentum space can be written as
\begin{align}
\mathcal{H}(\bm{q}) &= m_z \sigma_z  -4 t_0 \cos(q_x a)  \cos(q_y a) \otimes \bm{1} + 2t_{\rm so}^{(0)} \sin (q_xa) \sigma_x + 2t_{\rm so}^{(0)} \sin (q_ya)  \sigma_y ,
\end{align}
where $a=|\bm{N}_i|$ is the lattice constant representing the distance between the nearest neighbor sites. Expanding the Hamiltonian around the $\Gamma$ point yields 
\begin{align}
\mathcal{H}(\bm{q}) = 4 t_0 \bm{q}^2 + 2 t_{\rm so}^{(0)} (q_x \sigma_x + q_y \sigma_y ) + m_z  \sigma_z,
\end{align}
which renders the expected 2D Rashba type SO coupling. To generate 3D weyl type SO coupling, we simply remove the confining potential along $z$ direction and the tight binding model can be written as
\begin{align}
\mathcal{H}_{\rm 3D}(\bm{q}) &= \frac{p^2_z}{2m} - 4 t_0 \cos(q_x a)  \cos(q_y a)  + (m_z + \lambda_z p_z) \sigma_z  + 2t_{\rm so}^{(0)} \sin (q_xa) \sigma_x + 2t_{\rm so}^{(0)} \sin (q_ya)  \sigma_y.
\end{align}

\subsection{E. Realization for $^{133}$Cs Atoms}

\begin{figure}
\centering  \includegraphics[scale=0.3]{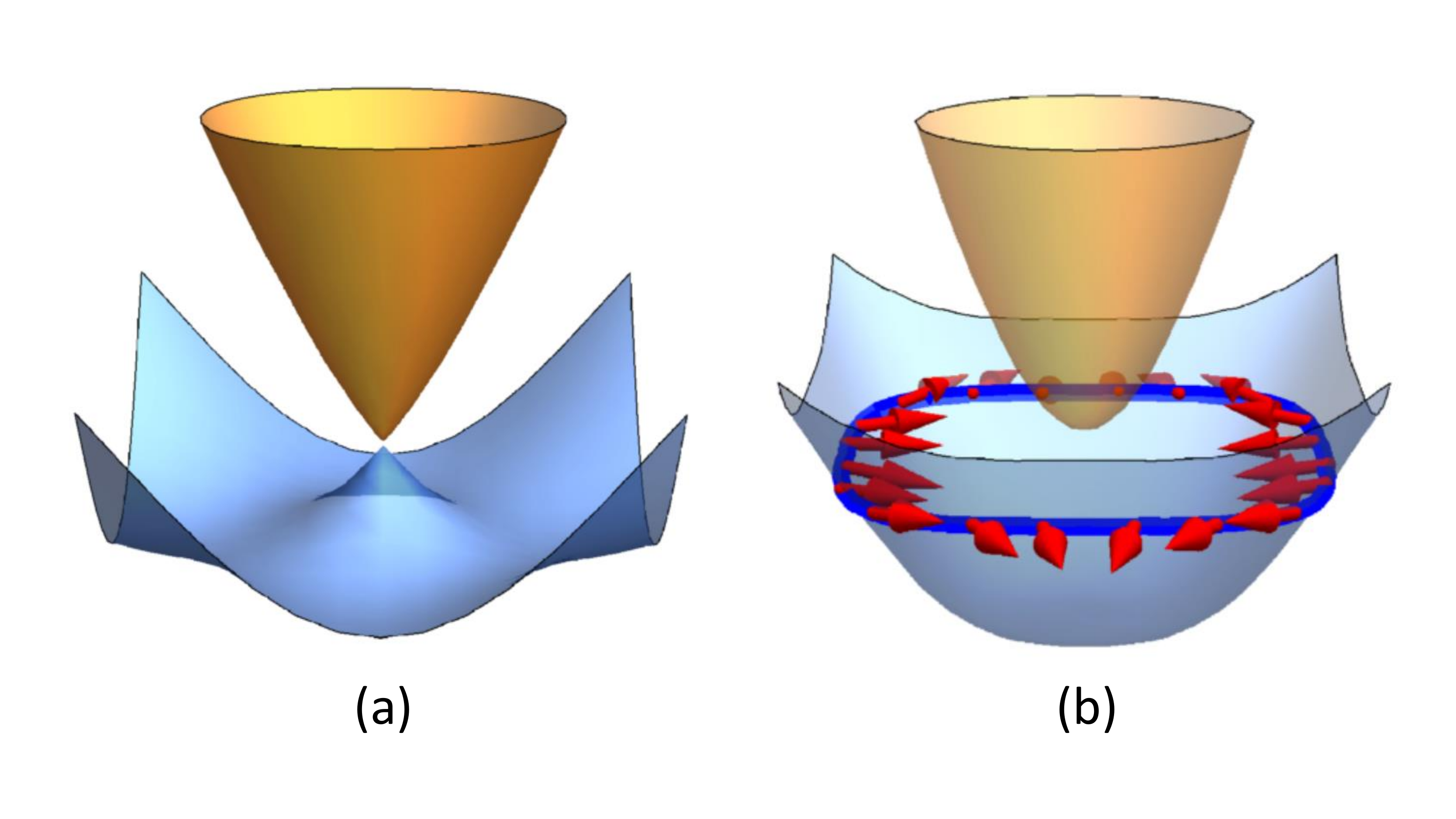}
\caption{Numerical results for $^{133}$Cs atoms. (a) The lowest two bands for the realized Hamiltonian around the $\Gamma$ point with $V_0=4E_r$, $M_0=1E_r$ and $m_z=0.68E_r$. (b) The spin texture of the state with energy $E=0$ for the Rashba SO coupled Hamiltonian with the parameters $V_0=4E_r$, $M_0=0.5E_r$ and $m_z=0.7E_r$, and spins are confined in $q_x$-$q_y$ plane.}
\label{realizationForCs}
\end{figure}

As discussed above, we choose the state $|4,-4 \rangle$ from the ground state as spin up $|g_{\uparrow} \rangle$, and $|3,-3 \rangle$ as spin down $|g_{\downarrow} \rangle$. The lattice potential includes a small correction to the purely spin-dependent term and can be written as
\begin{eqnarray*}
V_{\rm latt} =  \left(V_0 \bm{\sigma_z} + \delta V_0 \bm{1} \right)   \left(  \sin 2k_0x +  \sin 2k_0y  \right),
\end{eqnarray*}
where $V_0 = \frac{7}{12} E_{\rm V}^2  \alpha^2_{D_1} \left( \frac{1}{\Delta_{3/2}} - \frac{1}{\Delta_{1/2}}  \right)$ and the correction $\delta V_0 \sim 0.1 V_0$. In the following we show that $\delta V_0$ can induce novel effects. To simplify the proceeding analysis, we denote the tight binding model and the hopping terms as
\begin{eqnarray*}
-4 (t_0  \bm{1} + \delta t_0 \otimes \bm{\sigma}_z) \cos(q_x a)  \cos(q_y a),
\end{eqnarray*}
where $t_0 = (t_{\uparrow} + t_{\downarrow})/2$ and $\delta t_0 = (t_{\uparrow} - t_{\downarrow})/2$. Obviously, there is $\delta t_0  \sim 0.1 t_0 $ in the tight binding model. Then we redefine the effect  Zeeman term $m^{\rm eff}_z$ as
\begin{eqnarray*}
m^{\rm eff}_z = m_z - 4\delta t_0 \cos(q_x a)  \cos(q_y a).
\end{eqnarray*}
A few novel effects are followed. First, at the $\Gamma$ point $m^{\rm eff}_z=m_z - 4\delta t_0$, which vanishes when $m_z = 4\delta t_0$, and the band crossing is obtained [Fig.~\ref{realizationForCs}(a)]. Nevertheless, the Zeeman splitting at other symmetric points $\bm k=(0,\pi),(\pi,0),(\pi,\pi)$ keeps, showing that the time-reversal symmetry is indeed broken with nonzero $m_z$ and $\delta t_0$. Secondly, in the case of a nonzero Zeeman splitting at $\Gamma$ point, one can tune $m_z$ properly so that the spins of states at a finite energy $E$ exactly point to $x-y$ plane, namly, at such energy $m^{\rm eff}_z (\bm{q})=0$ so that the polarizations for states at $\bm{k}$ and $-\bm{k}$ are opposite [Fig.~\ref{realizationForCs}(b)]. In such region the spin texture resembles the case without Zeeman splitting. This effect can have novel consequences. For bosons, one can tune that the purely in-plane spin texture results in the band bottom, leading to a strong spin uncertainty in the ground states. For fermions, one can tune that the purely in-plane spin texture is obtained around Fermi surface. In this case under an attractive interaction the superfluid pairing $\Delta = U_0 \langle \hat{c}_{\bm{k},\uparrow} \hat{c}_{-\bm{k},\downarrow} \rangle$ can be largely enhanced even the Zeeman splitting is present, being different from a purely Rashba system where the Zeeman splitting always polarizes the Fermi surface and reduces the superfluid pairing. The topological gap of such a superfluid can then be greatly enhanced. Finally, when $-4 \delta_0 < m_z < 4 \delta_0$, each single energy band is topological nontrivial with the Chern number ${\rm Ch}_1 = \pm 1$.

According to the former discussions, the Raman potential takes the form
\begin{align*}
M = 2M_0   e^{2ik_0z}  \left( \cos 2k_0x + i \cos 2k_0y \right).
\end{align*}
Note that the Raman light $\bm{E}_{R1}$ also forms a standing wave in $x$-$y$ plane and generate a spin-independent lattice $V_{\rm in} \propto \cos^2 2k_0x +  \cos^2 2k_0y$.  This potential has the same contribution to both spin-up and spin-down atoms. Hence, it does not change any result that we obtain above.

\subsection{F. Spin Texture Around The Weyl Point}

\begin{figure}
\centering \includegraphics[scale=0.3]{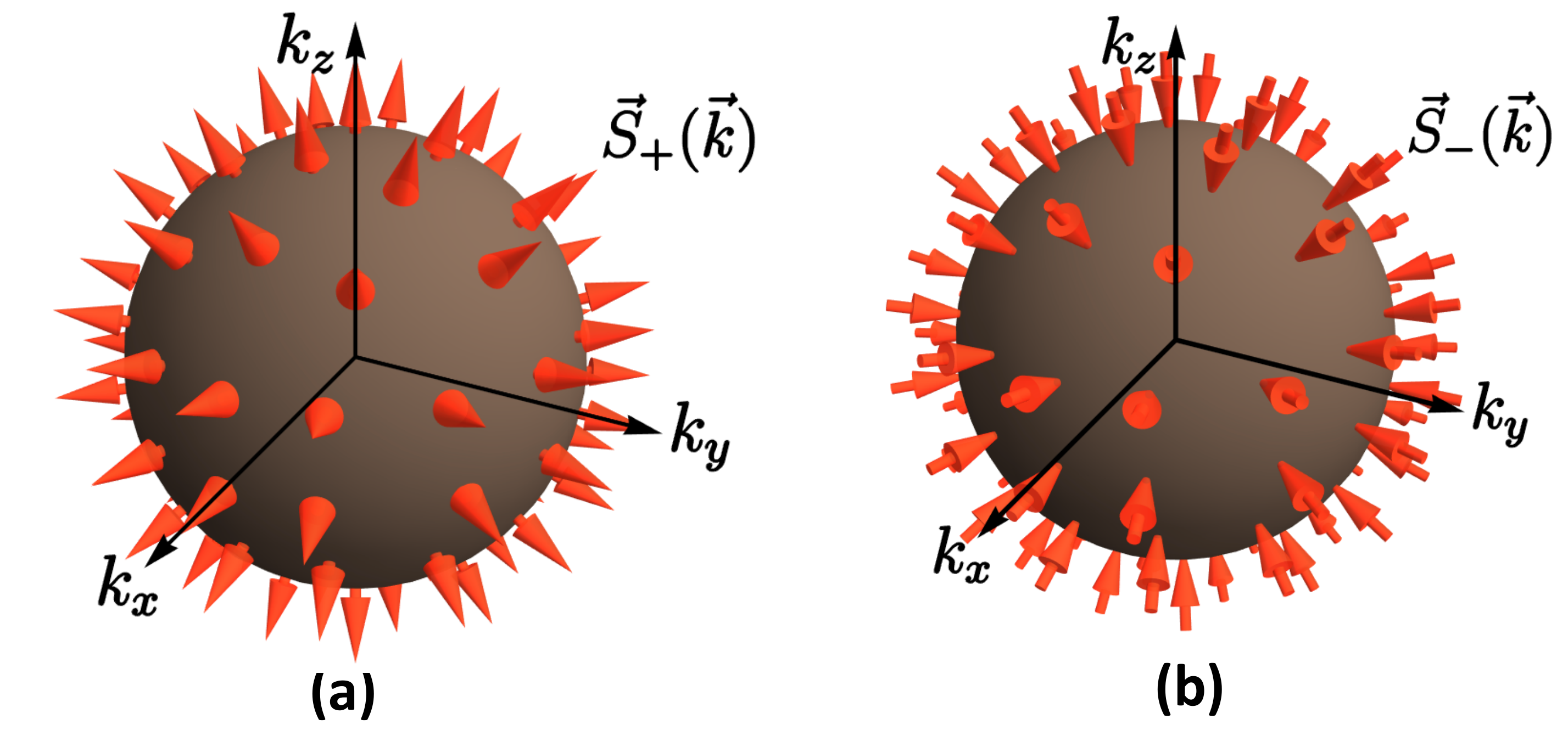}
\caption{The spin textures on the  two fixed-energy surfaces around the Weyl point. The Chern number defined on the two closed surfaces are $+1$(a) and $-1$(b), respectively. }
\label{spintexture}
\end{figure}

To simplify the proceeding analysis, we assume the Hamiltonian around the Weyl point is isotropic, which would not affect the topological property of the system. The Hamiltonian can be simplified as
\begin{align}
\mathcal{H}(\bm{q}) = \frac{\bm{q}^2}{2m_{\rm eff}} + v_F \bm{q} \cdot \bm{\sigma}.
\end{align}
The eigenvalues are $E_{\pm} = \frac{\bm{q}^2}{2m_{\rm eff}} \pm v_F |\bm{q}|$, and the degeneracy is the Weyl point at $\bm{q}=0$. The normalized eigenstates can be written as
\begin{align}
\bm{u}_{\pm}(\bm{q}) = \frac{1}{\sqrt{2q(q \pm q_z)}} \left(q_x - i q_y, - q_z \mp q \right).
\end{align}

We next consider the spin texture on the fixed-energy surfaces around the Weyl point. There are two closed spherical surfaces $F_{+}$ and $F_{-}$ with fixed energy in the 3D Brillouin zone. And the spin polarization on these surfaces can be defined as $\bm{S}_{\pm} (\bm{q}) = \langle \bm{u}_{\pm}(\bm{q}) | \bm{\hat{\sigma}} |\bm{u}_{\pm}(\bm{q}) \rangle.$ By substituting the equation (S20) into the defination, we obtain a simple result
\begin{align*}
\bm{S}_{\pm} (\bm{q}) = \pm \frac{\bm{q}}{|\bm{q}|}.
\end{align*}

The spin textures on the  two spherical surfaces $F_{+}$ and $F_{-}$ are drawn in the Fig.~\ref{spintexture}. And we can define the first Chern number  on the 2D spherical surface to characterize the spin texture
\begin{align*}
Ch_1 = \frac{1}{4\pi} \int_{F_{\pm}} \bm{S} \cdot \left( \partial_1 \bm{S} \times \partial_2 \bm{S} \right) \ \diff^2 \bm{k}.
\end{align*}
By employing spherical coordinates and parametrize the vector $\bm{S}_{\pm} (\bm{q}) = \pm \left( \sin \theta \cos \phi, \sin \theta \sin \phi, \cos \theta \right)$, the Berry curvature $\Omega_{\pm}$ can be easily caculated
\begin{align*}
\Omega_{\pm} = \bm{S}_{\pm} \cdot \left( \frac{\partial_{\theta}}{q} \bm{S}_{\pm} \times \frac{\partial_{\phi}}{q \sin \theta} \bm{S}_{\pm} \right)  = \pm \frac{1}{q^2}.
\end{align*}
We immediately find in the following for the Berry field strength
\begin{align*}
\bm{V} = \pm \frac{\bm{q}}{q^3}.
\end{align*}
This field looks like a magnetic field which is generated by a monopole at the Weyl point $\bm{q}=0$ with the strength $\pm 1$. Hence the Weyl node is a monopole or antimonopole for Berry curvature. Two fixed-energy surfaces $F_{+}$ and $F_{-}$ always contain the Weyl node with the strength $+1$ and $-1$. Hence the Chern number defined on the sphere $F_{+}$ and $F_{-}$ would be $Ch_{1} = 1$ and $Ch_{1} = -1$.

\section*{III. Estimation of the Lifetime}

Heating by the lattice and Raman lights is mainly induced by the spontaneous scattering of photons, which is random and causes fluctuations of the radiation force. For alkali atoms, a general expression of the photon scattering rate $\Gamma_{\rm sc}$ for the $D$ line doublet $^{2}S_{1/2} \to ^{2}P_{1/2}, ^{2}P_{3/2}$~\cite{grimm2000optical} can be written as
\begin{align}
\Gamma_{\rm sc} (\bm{r}) = \frac{\pi c^2 \Gamma^2}{2 \hbar \omega_0^3} \left( \frac{2+\mathcal{P} g_F m_F}{\Delta^2_{D_2}} + \frac{1 - \mathcal{P} g_F m_F}{\Delta^2_{D_1}}  \right) I(\bm{r}),
\end{align}
where $g_F$ is the Land\'e factor, $\mathcal{P}$ denotes the laser polarization ($\mathcal{P}=0, \pm 1$ for linearly and circularly $\sigma^{\pm}$ polarized light), $\omega_0$ is the averaged resonant frequency of the $D$ line doublet, and $I(\bm{r})$ is the laser beam intensity. In experiment, we assume that the atoms stay at the center of a laser beam $\bm{E}$, and then the laser beam intensity can be written as $I(\bm{r}) = 1/2 \epsilon_0 c |\bm{E}|^2$. With the scattering rate, we can estimate the heating rate $\dot{T}$ by
\begin{align}
k_{\rm B} \dot{T} = \frac 23 E_r \Gamma_{\rm sc}.
\end{align}
And then the lifetime can be estimated as $\tau \sim 100 \nano \kelvin/\dot{T}$, where $100\nano \kelvin$ is the typical difference between the initial experimental temperature and the critical temperature.

For the scheme of Dirac type SO couping, we consider the realization for $^{87}$Rb Bose-Einstein condensate (BEC) and $^{40}$K degenerate Fermi gases. For the $^{87}$Rb atoms, the dipole trap is formed by two far-detuned laser beams and the estimated heating rate is about $10 \nano \kelvin/\second$~\cite{wu2016realization}. The lifetime with the dipole trap alone is about 10 seconds. Then we consider the heating from lattice and Raman beams, which are applied with the wavelength $\lambda = 786 \micro \metre$  and the beam waist $\varpi = 200 \micro \metre$. Under typical parameter conditions $V_0 = 4E_r$ and $M_0 = 1E_r$, we find  the scattering rate $\Gamma_{\rm sc} \approx 0.54 \hertz$. Since the recoil energy $E_r/\hbar = 2\pi \times 3707 \hertz$, the total heating rate of the lattice and Raman beams is about $\dot{T}=63.7\nano \kelvin /\second$. Therefore, the lifetime of $^{87}$Rb BEC is about $\tau \sim \frac{100\nano\kelvin}{73.7\nano\kelvin/\second} \approx 1.36 \second$. For the $^{40}$K degenerate fermion gas, to realize typical parameter conditions $V_0 = 4E_r$ and $M_0 = 1E_r$, we find the heating rate of lattice and Raman lights with wavelength $\lambda = 768\nano \metre$ is about $665 \nano \kelvin/\second$, which gives the lifetime about $\tau \sim 150 \milli \second$. If considering a smaller lattice, one can even enhance the lifetime up to several hundreds of milliseconds.

For the scheme of Rashba and Weyl type SO coupling, the application for $^{133}$Cs atoms has been considered, and the lifetime of $^{133}$Cs atoms can be even much longer. The lifetime with only the dipole trap is approximately $15\second$~\cite{weber2003bose}. The lattice potential is generated by the laser beam with the wavelength $\lambda = 916.6 \nano \meter$, which couples the gound state and $6^2P_{1/2}$($6^2P_{3/2}$) levels. To produce the lattice potential with a typically depth  $V_0 = 4E_r$, we apply the laser beam with the waist $\varpi = 200 \micro \meter$ and the power $P = 15.4 \milli \watt$. And then the heating rate of the lattice light can be estimated as $3\nano\kelvin/\second$, which is very small because the detuning $\Delta$ is very large. Then we consider the heating rate of the Raman lights. The Raman potental is generated by two light beams with the wavelength $\lambda = 458.3 \nano \meter$, which couples the ground state and $7^2P_{1/2}$($7^2P_{3/2}$) levels. To produce the Raman potentials with a typically strength $M_0 = 1E_r$, we apply the laser beams with the waist  $\varpi = 200 \micro \meter$ and the power $P = 21 \milli \watt$. The heating rate of the Raman lights is also considerably small about $10\nano \kelvin / \second$. With these results we can estimate the lifetime of the $^{133}$Cs atoms to be $\tau >5\second $.


\end{document}